\documentclass[]{aa}  

\usepackage{graphicx}
\usepackage{txfonts}
\usepackage{booktabs}

\usepackage[]{hyperref}
\begin{document} 

    \title{Reconstructing the Sun’s Alfvén surface and wind braking torque with Parker Solar Probe}

   \titlerunning{Reconstructing the Sun’s Alfvén surface}
   \authorrunning{Finley et al.}

   \author{A. J. Finley\inst{1}
          }

    \institute{Universit\'e Paris-Saclay, Universit\'e Paris Cit\'e, CEA, CNRS, AIM, 91191, Gif-sur-Yvette, France\\ \email{adam.finley@cea.fr} }

   \date{Received July 30, 2025; accepted -- --, 202-}

\abstract{The Alfvén surface -- where the solar wind exceeds the local Alfvén speed as it expands into interplanetary space -- is now routinely probed by NASA's Parker Solar Probe (PSP) in the near-Sun environment. The size of the Alfvén surface governs how efficiently the solar wind braking torque causes the Sun to spin-down.}
{We aimed to characterise the size and evolution of the Alfvén surface as magnetic activity increased during solar cycle 25.}
{The Alfvén surface was extrapolated from the solar wind mass and magnetic flux measured by the SWEAP and FIELDS instrument suites onboard PSP. We accounted for the acceleration of the solar wind along Parker spiral magnetic field lines and used potential field source surface modelling to determine the sources of the solar wind.}
{The longitudinally averaged Alfvén radius measured by PSP grew from 11 to 16 solar radii as solar activity increased. Accordingly, the solar wind angular momentum-loss rate grew from $\sim$1.4$\times 10^{30}$ erg to 3$\times 10^{30}$ erg. Both the radial and longitudinal scans of the solar wind contained fluctuations of 10-40\% from the average Alfvén radius in each encounter. Structure in the solar corona influenced the morphology of the Alfvén surface, which was smallest around the heliospheric current sheet and pseudostreamers.}
{The Alfvén surface was highly structured and time-varying however, at large-scales, organised by the coronal magnetic field. The evolution of the solar corona over the solar cycle systematically shifted the magnetic connectivity of PSP and influenced our perception of the Alfvén surface. The Alfvén surface was 30\% larger than both thermally-driven and Alfvén wave-driven wind simulations with the same mass-loss rate and open magnetic flux, but had a similar dependence on the wind magnetisation parameter.}


   \maketitle
%

\section{Introduction}

As the solar wind accelerates away from the Sun \citep{parker1958dynamics} it transitions from sub-Alfvénic to super-Alfvénic speeds \citep[exceeding the speed of plasma waves restored by magnetic tension;][]{alfven1942existence}. The divide between these two regimes is known as the Alfvén surface and is located between 5 to 20 solar radii in global solar wind simulations \citep[e.g.][]{chhiber2019contextual, reville2020role}. This transition may play an important role in the turbulent heating of the solar wind and the formation of magnetic ``switchbacks'' \citep[e.g.][]{ruffolo2020shear}. In the sub-Alfvénic regime, disturbances in the solar wind produce waves that can propagate back towards the Sun \citep{belcher1971large} and establish a causal feedback on the outflow \citep{verdini2009turbulence, chandran2021approximate}. The extent of the sub-Alfvénic region around the Sun affects the strength of magnetic torques that enforce rotation in the solar corona \citep{kasper2019alfvenic}, the amount of angular momentum lost in the solar wind \citep{weber1967angular, marsch1984distribution}, and the subsequent spin-down of the Sun \citep{finley2018effect}. The nature of this division, either a sharp surface or turbulent transition \citep[e.g.][]{chhiber2022extended}, could also impact the search for star-planet magnetic interactions \citep{saur2013magnetic}; analogous to the magnetic interactions between Jupiter and its moons \citep{clarke2004jupiter}. 

By extrapolating in-situ measurements of the solar wind from 1au back to the Sun, previous authors have shown that the Alfvén surface expands and contracts over the solar cycle \citep{goelzer2014analysis, kasper2019strong}. Estimates from in-situ measurements place the Alfvén surface between 10 to 40 solar radii \citep[e.g.][]{verscharen2021flux}. The addition of remote-sensing observations narrows this range towards 10 to 20 solar radii \citep{deforest2014inbound, telloni2021exploring}. NASA's Parker Solar Probe (PSP) has been exploring the near-Sun environment since its launch in 2018 \citep{fox2016solar}. PSP began to enter the sub-Alfvénic solar wind as its perihelion distance decreased and solar activity increased \citep[the first time was in encounter 8;][]{kasper2021parker}. Due to its proximity to the Sun, extrapolating the distance of the Alfvén surface from PSP reduces the associated uncertainties. Using PSP, both \citet{cranmer2023sun} and \citet{mann2023heliospheric} estimated that the Alfvén surface lies between 8 to 19 solar radii. 

\begin{figure*}
    \centering
    \includegraphics[trim=0cm 0cm 0cm 0cm, clip, width=\textwidth]{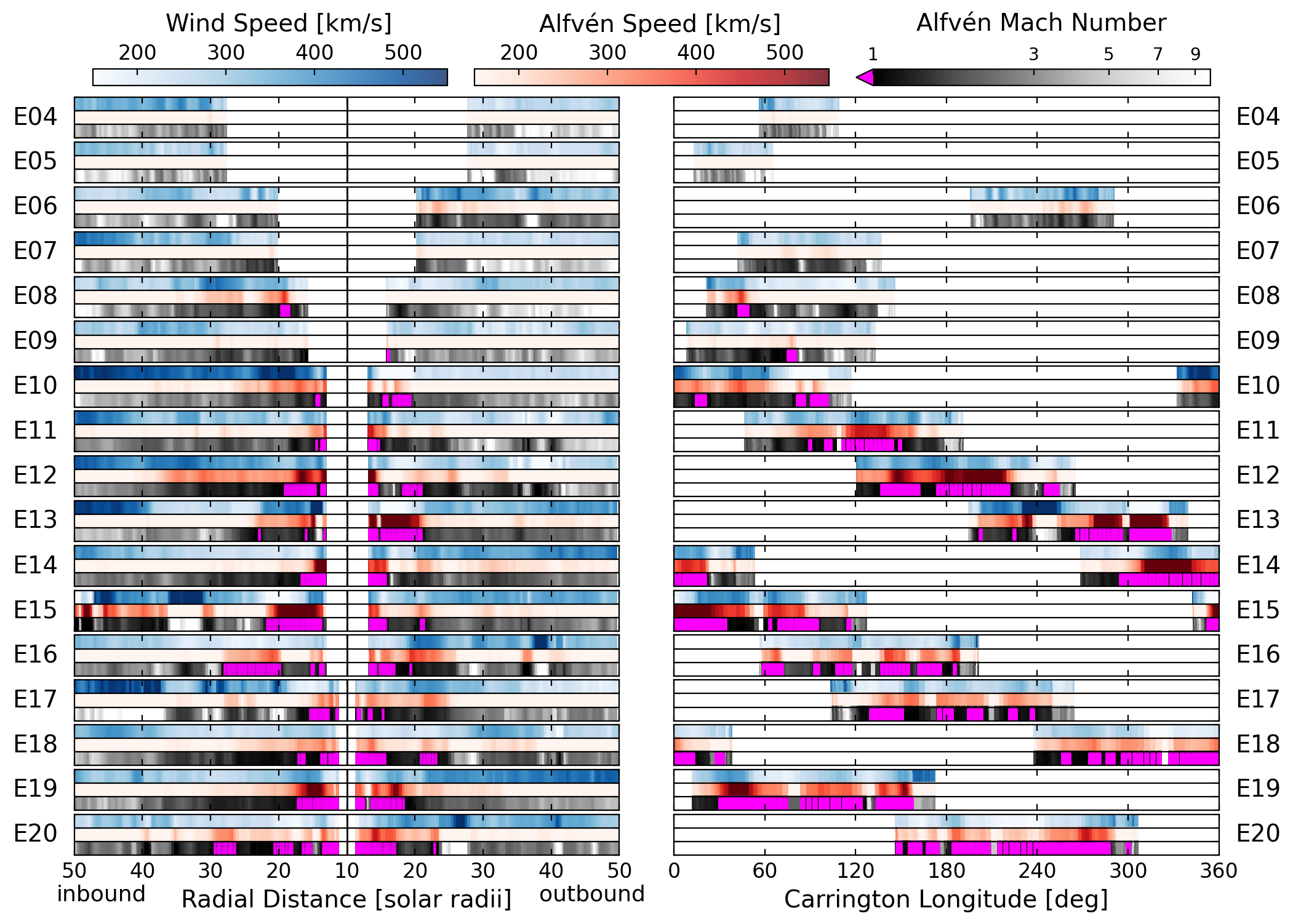}
    \caption{Summary of the hourly solar wind speed, Alfvén speed, and Alfvén Mach number measured by PSP from E04 to E20. Intervals containing sub-Alfvénic solar wind are highlighted in magenta. Left: Quantities displayed as a function of PSP's radial distance. PSP's inbound journey is on the left, the outbound journey on the right. Structure in the radial scans is quickly discerned in this view, but the longitudinal scans are compressed. Right: Quantities shown as a function of PSP's Carrington longitude. The radial scans are obscured, but the structure in each longitudinal scan is easily distinguished.}
    \label{fig:speedmachsummary}
\end{figure*}

Despite these advances, the structure and long-term evolution of the Sun's Alfvén surface remains uncertain. In this study, we estimate the shape and size of the Alfvén surface in the ecliptic plane using in-situ measurements from PSP. Section 2 assembles the observations from PSP and describes the method used to estimate the size of the Alfvén surface. Section 3 summarises the general trend of the Alfvén surface expanding with rising solar activity. Section 4 assesses the spatial and temporal variation of the Alfvén surface using the longitudinal and radial scans in each PSP encounter. Section 5 connects the shape of the Alfvén surface to the structure of the underlying coronal magnetic field. Section 6 compares the extent of the Alfvén surface with magnetohydrodynamic simulations of the Sun's angular momentum-loss rate. A more detailed analysis, connecting the Alfvén surface with solar wind turbulence and rotational flows, is left for future work.

\section{Alfvén surface reconstruction}

Following a sequence of gravitational assists with Venus, PSP's perihelion distance decreased from 35.7 solar radii in 2018 (encounters E01 to E03) to 27.9 solar radii in 2019 (E04 and E05), 20.4 solar radii in 2020 (E06 and E07), 16.0 solar radii in 2021 (E08 to E09), 13.3 solar radii in late 2021 (E10 to E16), 11.4 solar radii in 2023 (E17 to E21) and finally to 9.9 solar radii in late 2024 (E22 to present). This produced progressively longer encounters (time spent below 30 solar radii) that scanned over larger ranges of Carrington longitudes (from $\sim$ 10$^{\circ}$ to 200$^{\circ}$). Accordingly, PSP's speed at closest approach increased from around 90 to 190 km/s which shifted the apparent solar wind direction with respect to the onboard instruments.

\begin{figure*}
    \centering
    \includegraphics[trim=0cm 0cm 0cm 0cm, clip, width=0.95\textwidth]{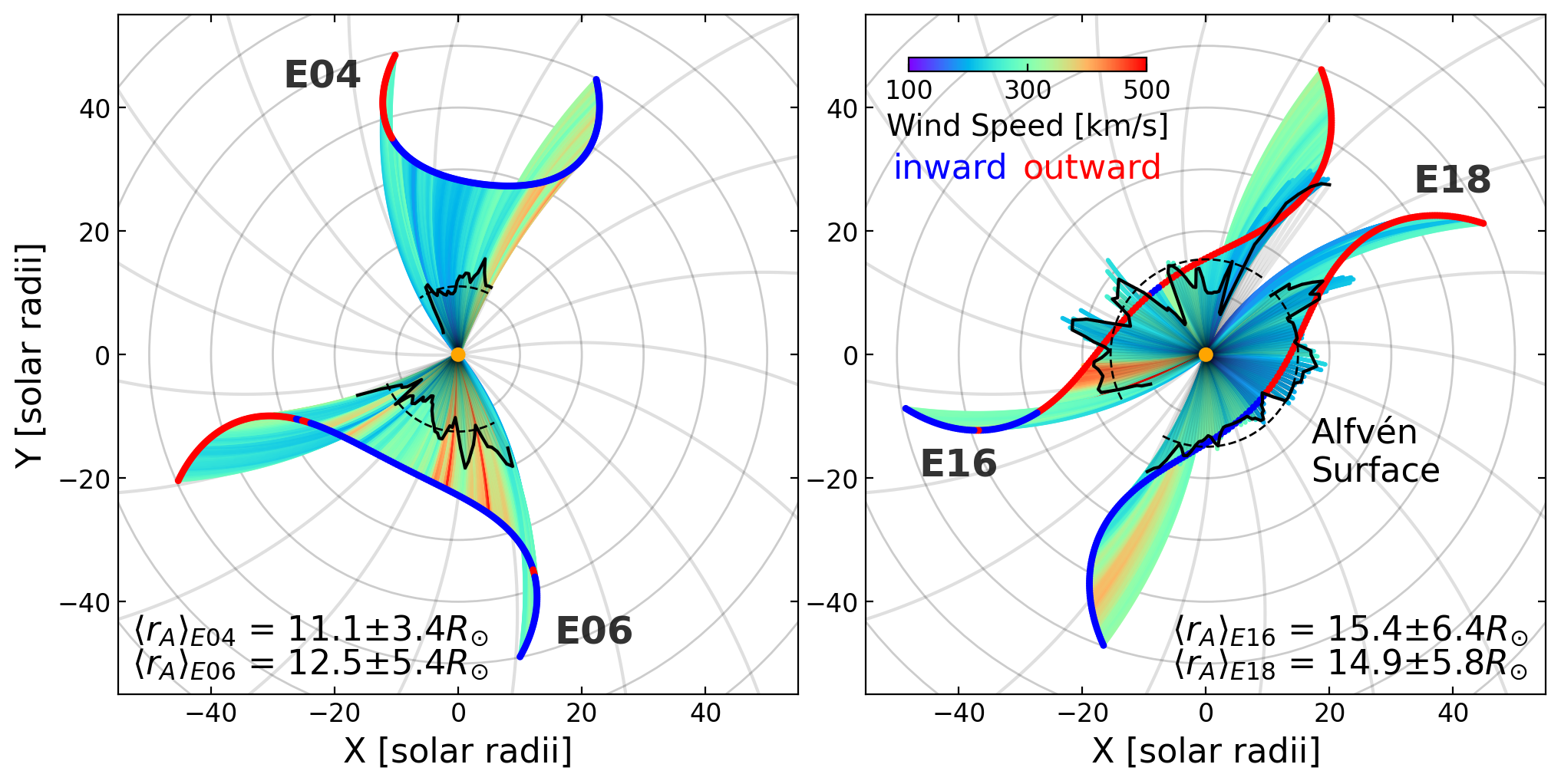}
    \caption{Extrapolation of the solar wind along Parker spiral magnetic field lines for PSP encounters E04 and E06 (left), and E16 and E18 (right). The path of PSP is coloured red or blue according to the radial magnetic field polarity. Parker spiral magnetic field lines are coloured by solar wind speed using an acceleration of $\alpha=0.1$. The solid black lines show the Alfvén surface when averaged into bins of two degrees width in Carrington longitude. The dashed semi-circles represents the longitudinally averaged Alfvén radii. }
    \label{fig:exampleOrbits}
\end{figure*}

\begin{figure}
    \centering
    \includegraphics[trim=0cm 0cm 0cm 0cm, clip, width=0.48\textwidth]{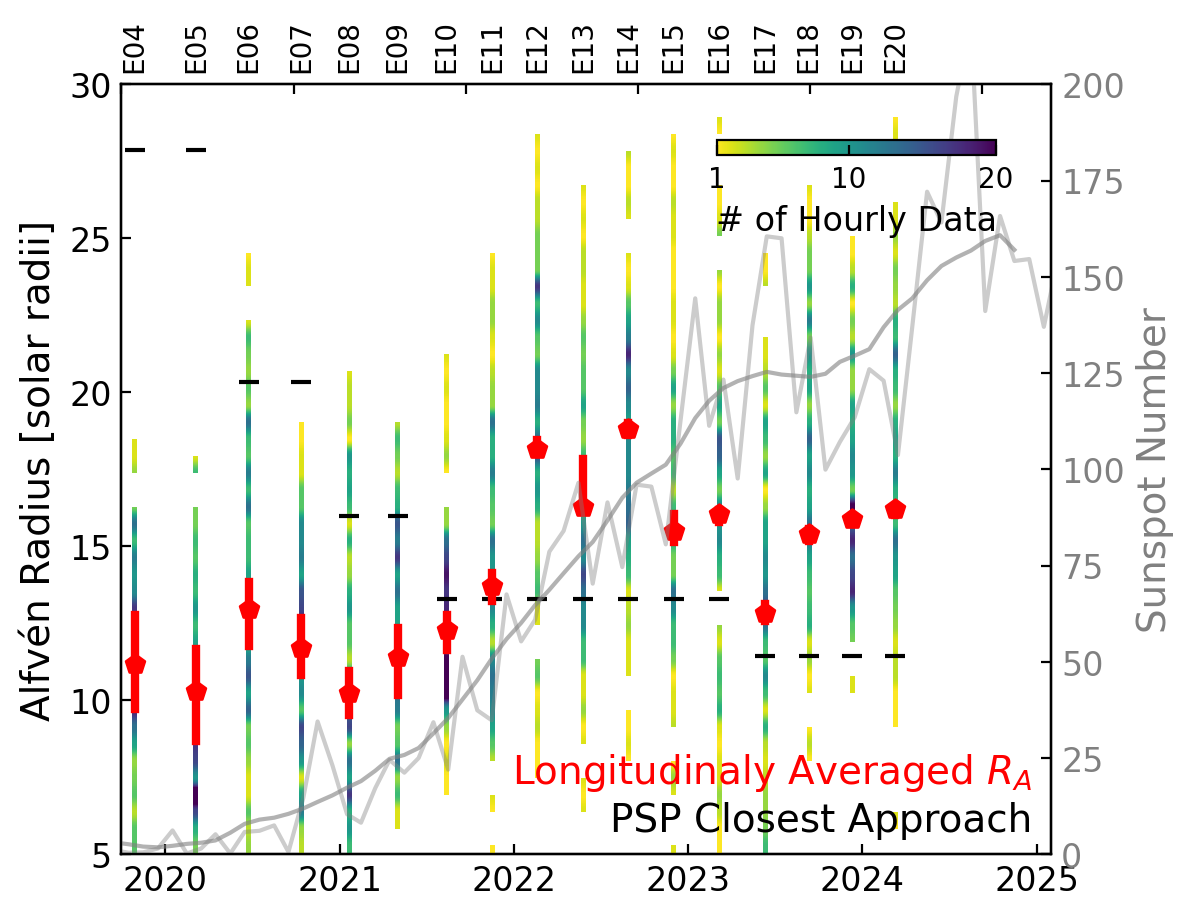}
    \caption{Evolution of the Alfvén radius extrapolated from PSP over solar cycle 25. The horizontal black ticks show PSP's closest approach in each encounter. The distribution of hourly extrapolations using $\alpha=0.1$ are shown with coloured histograms. The longitudinally averaged value for each encounter is indicated with a red marker. The associated red bars show the variation in the average due to different wind accelerations (from $\alpha=0$ to $\alpha=0.2$). The thin and thick grey lines represent the monthly and monthly smoothed sunspot number from WDC-SILSO.} 
    \label{fig:cycle}
\end{figure}

We compiled hourly averaged observations of the solar wind speed $v$, density $\rho$, and magnetic field strength $B$ from when PSP was within 50 solar radii of the Sun. An hourly timescale was chosen to remove small-scale fluctuations, such as magnetic switchbacks, in order to focus on the organisation of the solar wind at large-scales. We combined the solar wind speed measurements from the SPC \citep{case2020solar} and SPAN-I \citep{livi2022solar} instruments in the SWEAP suite; typically SPAN-I was used within 35 solar radii when the majority of the solar wind velocity distribution was visible behind PSP's heat shield \citep[see][for information relating to the design of the SWEAP suite]{kasper2016solar}. Similarly, we combined the proton density measurements from the two instruments but adopted densities from FIELDS quasi-thermal noise spectroscopy \citep{moncuquet2020first} when available. The magnetic field strength was taken from the fluxgate magnetometer (MAG) in the FIELDS suite \citep{bale2016fields}. The local Alfvén speed was calculated from the hourly averaged variables as 
\begin{equation}
    v_A = \frac{B}{\sqrt{\mu_0 \rho}}.
\end{equation}
In future, this could be improved by accounting for other particles species that carry a significant momentum, such as the alpha-particles \citep{finley2021contribution, mcmanus2022density}. Figure \ref{fig:speedmachsummary} summarises $v$, $v_A$, and the Alfvén Mach number $M_A=v/v_A$ from E04 to E20 as a function of PSP's distance from the Sun and its Carrington longitude, with the sub-Alfvénic intervals highlighted. A similar analysis was performed by \citet{chhiber2024alfven} who searched for sub-Alfvénic intervals as small as 10 minutes from E08 to E14.

From the in-situ measurements, we used the conservation of mass flux and magnetic flux along a steady-state, radially expanding flux tube 
\begin{equation}
    F_m = \rho(r) v(r) r^2 = \text{const.},
\end{equation}
\vspace{-0.5cm}
\begin{equation}
    F_B = B(r) r^2 = \text{const.},
\end{equation}
where $r$ is the radial distance of PSP, to extrapolate the radial variation of density $\rho(r)$ and magnetic field $B(r)$. Appendix \ref{ap:summary} contains a summary of $F_m$ and $F_B$ in the same style as Figure \ref{fig:speedmachsummary}. The turbulent and time-varying nature of the solar wind are the primary limitations of these equations. We accounted for a radial acceleration of the solar wind with a power-law function,
\begin{equation}
    v(r) = v_{in}\bigg(\frac{r}{r_{in}}\bigg)^{\alpha},
\end{equation}
where $\alpha$ is the exponent, and the subscript $in$ denotes in-situ measurements like the solar wind speed $v_{in}$ and the radial distance of each measurement $r_{in}$. The case of $\alpha=0$ corresponds to zero acceleration i.e. a constant wind speed of $v_{in}$. Based on the observed trends in solar wind speed versus distance \citep{schwenn1983average, schwenn1990large, venzmer2018solar}, we adopted a value of $\alpha=0.1$. 

From the equations above, we calculated the Alfvén Mach number versus radial distance
\begin{equation}
    M_A(r) = \frac{v(r)}{v_A(r)} =  \frac{v(r)}{B(r)}\sqrt{\mu_0 \rho(r)} = \frac{v_{in}}{B_{in}}\sqrt{\mu_0\rho_{in}}\bigg(\frac{r}{r_{in}}\bigg)^{\alpha/2+1} ,
\end{equation}
in order to locate the Alfvén radius $r_A$ where $M_A(r_A)=1$. The travel time of the solar wind between the Alfvén surface and PSP is
\begin{equation}
    \Delta t = \int_{r_A}^{r_{in}} \frac{dr}{v(r)}=  \frac{ r_{\text{in}}}{(1-\alpha)v_{\text{in}}}\bigg[ 1- \bigg(\frac{r_A}{r_{in}}\bigg)^{1-\alpha} \bigg],
\end{equation}
where a negative $\Delta t$ corresponds to the case when PSP is inside the Alfvén surface, i.e. $r_A>r_{\text{in}}$. The Sun's rotation twists the interplanetary magnetic field and forms the Parker spiral \citep{parker1958dynamics, neugebauer1998spatial}. Adopting the Sun's equatorial rotation rate of 24.5 days ($\Omega_{\odot}=14.7$ deg/days), the Carrington longitude where the solar wind crossed the Alfvén radius is
\begin{equation}
    \phi(r) = \phi_{in} - \Delta t \Omega_{\odot}.
\end{equation}

\begin{figure}
    \centering
    \includegraphics[trim=0cm 0cm 0cm 0cm, clip, width=.48\textwidth]{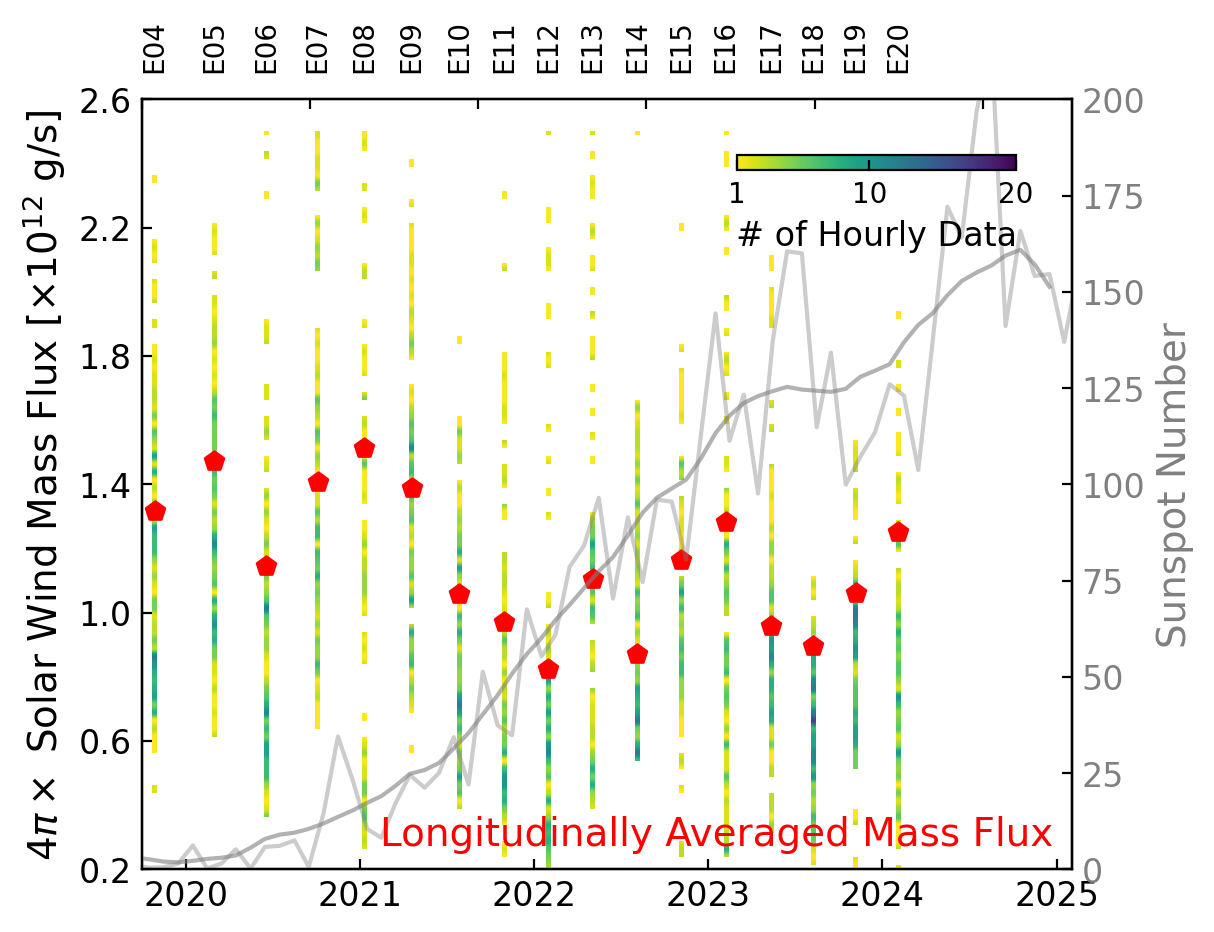}\\
    \includegraphics[trim=0cm 0cm 0cm 0cm, clip, width=.48\textwidth]{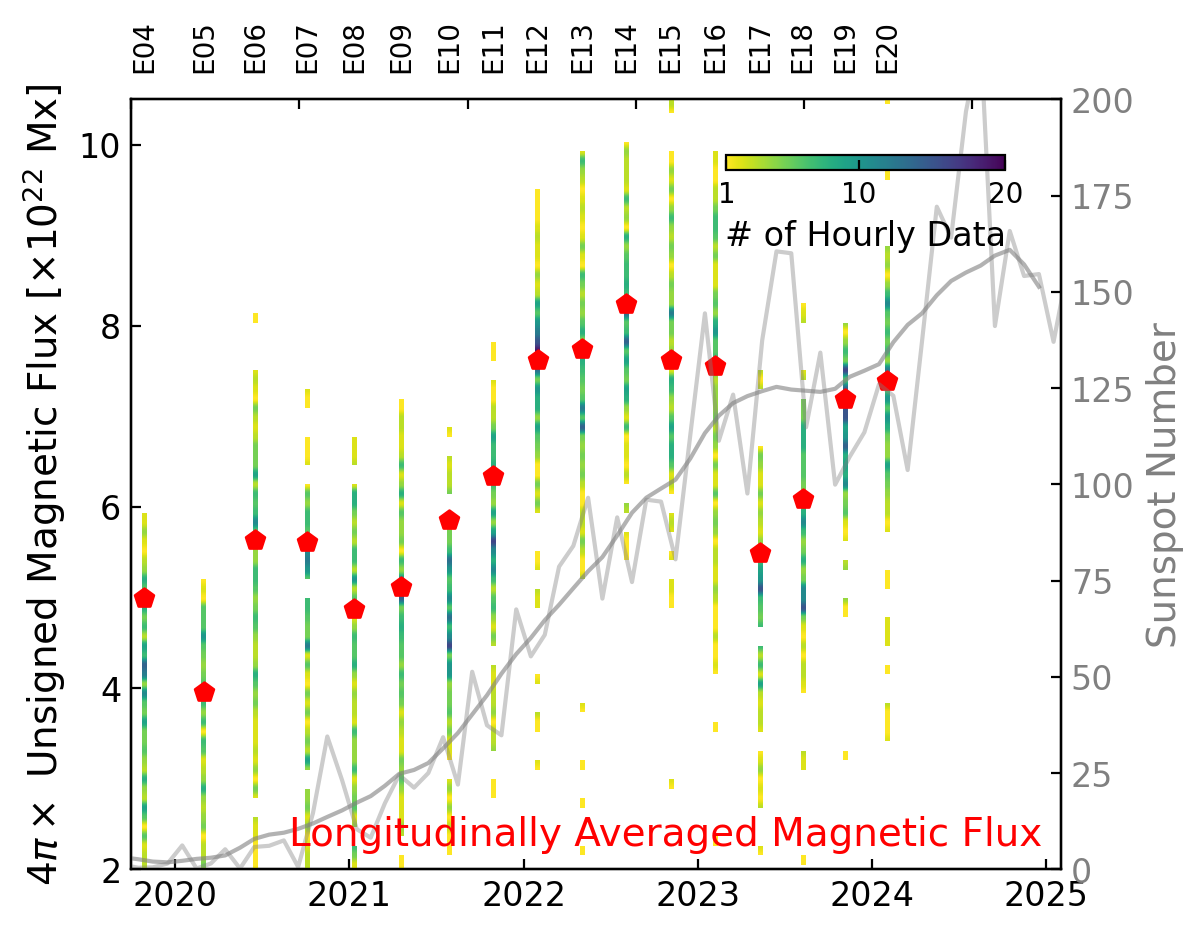}
    \caption{Same as Figure \ref{fig:cycle}, but now for the solar wind mass flux and magnetic flux. Both values have been multiplied by a factor of $4\pi$ to be used as an estimate for the global mass-loss rate and open magnetic flux.}
    \label{fig:massmassgcycle}
\end{figure}

We used the in-situ measurements from PSP to recover the shape and size of the Alfvén surface from E04 to E20; spanning the rising phase of solar cycle 25. Examples from E04, E06, E16, and E18 are shown in Figure \ref{fig:exampleOrbits} \citep[a similar approach was used in][for E10]{badman2023prediction}. PSP first crossed into sub-Alfvénic wind during E08, visible in Figure \ref{fig:speedmachsummary}\footnote{As we used hourly averaged data, Figure \ref{fig:speedmachsummary} does not highlight any of the sub-Alfvénic intervals with shorter durations.}, when the Alfvén speed dropped entering a low density wind stream. After this, PSP spent a growing fraction of each encounter in the sub-Alfvénic solar wind, from a few hours in E11 to over a day in E20, with the number of Alfvén surface crossings growing from 2 to 10. This caused the average distance between PSP and the Alfvén surface to decrease from $\sim$15-20 solar radii in E04 to $\sim$5 solar radii in E20, reducing the uncertainty in the extrapolations.

\section{Solar cycle variation in the average Alfvén radius}

We compiled the hourly estimates of the Alfvén radius in Figure \ref{fig:cycle} with comparison to the minimum distance of PSP and the progression of solar cycle 25 using the monthly sunspot number from the World Data Center SILSO\footnote{Data accessed May 2025: https://www.sidc.be/SILSO/cyclesmm} at the Royal Observatory of Belgium. The distribution of Alfvén radii evolved over the solar cycle with an overall trend of larger values at solar maximum than at solar minimum. However, the largest values for the Alfvén radius were obtained during the rising phase of the solar cycle. In order to meaningfully compare the Alfvén radius between encounters, we averaged the observations in longitudinal bins of two degrees width. Examples of this are shown in Figure \ref{fig:exampleOrbits}. From the binned surface we computed a longitudinally averaged Alfvén radius for each encounter. This was intended to reduce the bias from each encounter spending different amounts of time over certain Carrington longitudes (e.g. during the inbound and outbound radial scans).

From the longitudinally averaged Alfvén radius, we found an expansion of $\sim$5 solar radii from 2020 to 2024, from 11 solar radii at solar minimum up to 16 solar radii at solar maximum. The largest average Alfvén radius of 19 solar radii was found during the rising phase of the cycle. During the solar cycle, PSP's perihelion decreased and crossed below the average Alfvén radius during E10 and E11. The effect of varying the solar wind acceleration from $\alpha=0$ to 0.2 on the average Alfvén radius in the extrapolations is shown in Figure \ref{fig:cycle}. The wind acceleration had a larger effect when PSP was further from the Sun during the earlier encounters (with perihelia above 15 solar radii). From E04 to E10, changing the wind acceleration systematically shifted the Alfvén surface by 2 solar radii. This does not affect our conclusion that PSP measured an expanding equatorial Alfvén surface as solar activity increased.

The size of the Alfvén radius is intrinsically linked to the solar wind mass-loss rate and open magnetic flux \citep{reville2015effect}; further discussed in Section 6. Examining $F_m$ and $F_B$ as a function of solar cycle in Figure \ref{fig:massmassgcycle}, we find that while the mass-loss rate, $\dot{M} = 4\pi \langle F_m \rangle$, decreased slightly with rising solar activity (discussed in Appendix \ref{ap:summary}), the open magnetic flux, $\phi_{\text{open}} = 4\pi \langle |F_B| \rangle$, peaked during the rising phase of the cycle. The peak in open magnetic flux is similarly visible in the Alfvén radius in Figure \ref{fig:cycle}. During this time, however, the Sun's unsigned photospheric magnetic flux steadily increased without an obvious peak in 2022. From Figure \ref{fig:speedmachsummary} we identified that at this time (E12 to E15) PSP consistently observed stronger Alfvén speeds (>500km/s) in the longitudinal scans. Therefore, the increase in magnetic flux and size of the Alfvén radius may not have been a global effect and instead relate to local structures measured by PSP in those encounters.

\begin{figure*}
    \centering
    \includegraphics[trim=0cm 0cm 0cm 0cm, clip, width=\textwidth]{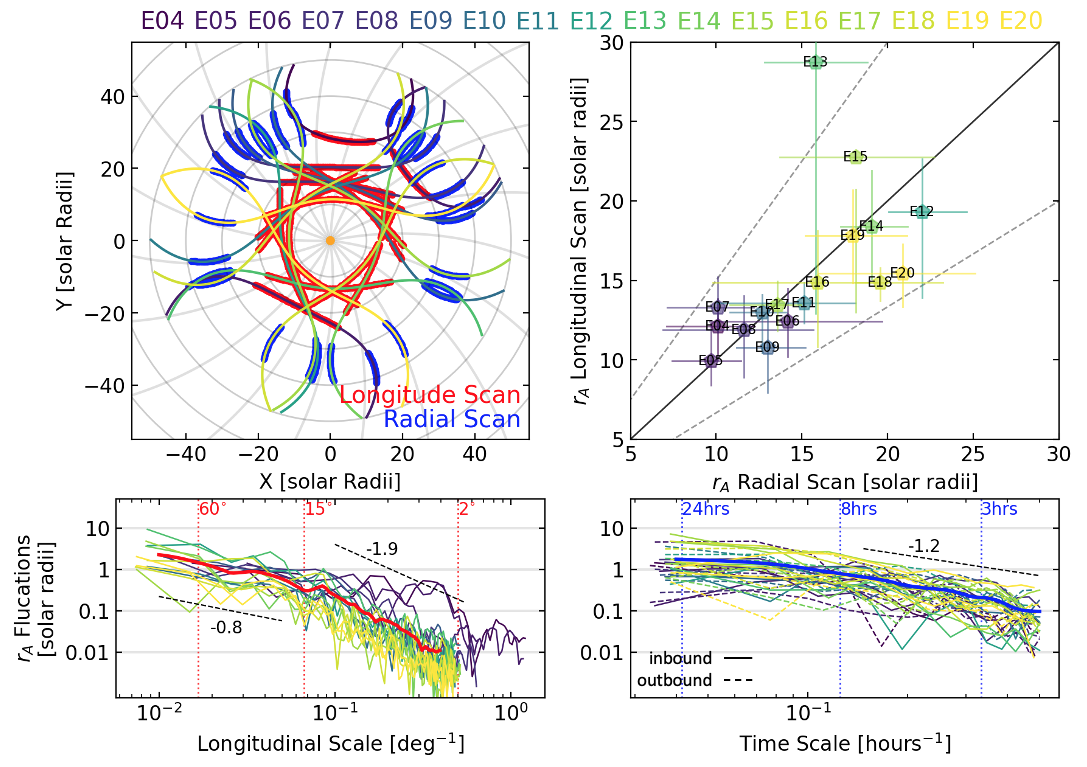}
    \caption{Fluctuations in the Sun's Alfvén radius during PSP's longitudinal and radial scans. Encounters are coloured by monthly sunspot number. Top left: PSP's trajectory in the Carrington frame from E04 to E20, when PSP was below 50 solar radii. The segments highlighted red at closest approach correspond to the longitudinal scans. The segments highlighted blue correspond to the radial scans, when PSP moved quasi radially in the Carrington frame. Top right: Markers, labelled by encounter, show the 50th percentile of the Alfvén radii measured during the longitudinal and radial scans. Vertical and horizontal lines represent the 25th to 75th percentiles during the longitudinal and radial scans, respectively. The solid line marks equality and the dashed lines highlight $\pm$ a third. The bottom panels show the longitudinal scan and radial scans decomposed with discrete Fourier transforms to identify the spatial and temporal scales associated with the variation of the Alfvén radius away from the average ($r_A-\langle r_A\rangle$). Thick solid lines show the averaged profile for all encounters except E04 and E05.}
    \label{fig:scanning}
\end{figure*}

\begin{figure}
    \centering
    \includegraphics[trim=0cm 0cm 0cm 0cm, clip, width=0.48\textwidth]{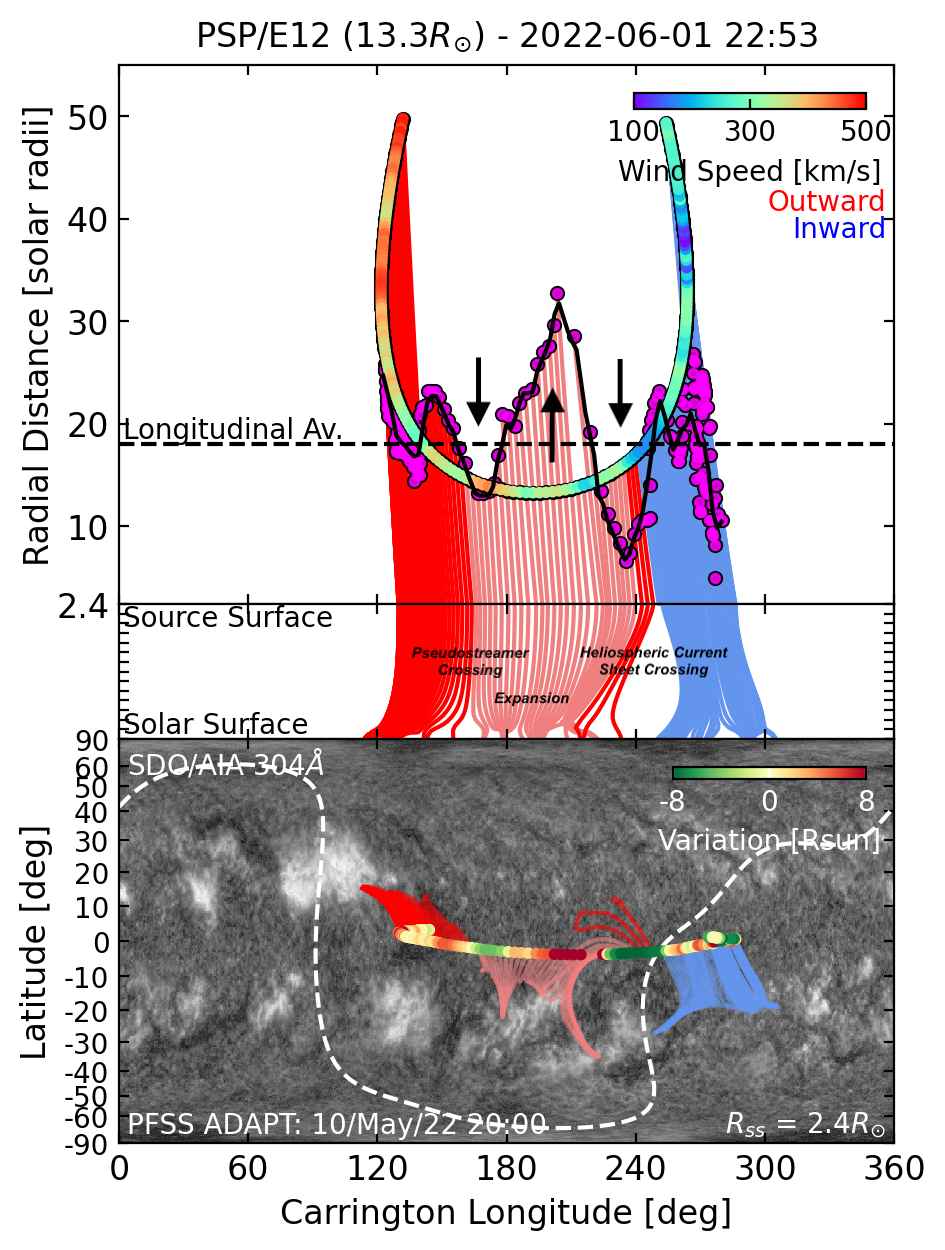}
    \caption{Mapping the solar wind from PSP to the Sun during E12 in June 2022. The top panel shows the solar wind speed at PSP with magnetic field lines traced down to the solar surface using a Parker spiral and PFSS model. Field lines connecting to the southern hemisphere are distinguished with a lighter colour. Magenta points show the Alfvén radius for each hourly extrapolation and the solid black line shows these points binned in longitude. The longitudinal average is highlighted with a dashed horizontal line. The bottom panel contains a Carrington map of SDO/AIA 304$\AA$ emission in greyscale with the heliospheric current sheet over-plotted with a white dashed line. Coloured markers show the variation of the Alfvén radius away from the longitudinal average. Magnetic field lines connect these points to their sources.}
    \label{fig:scanEncounter}
\end{figure}

\begin{figure*}
    \centering
    \includegraphics[trim=0cm 0cm 0cm 0cm, clip, width=\textwidth]{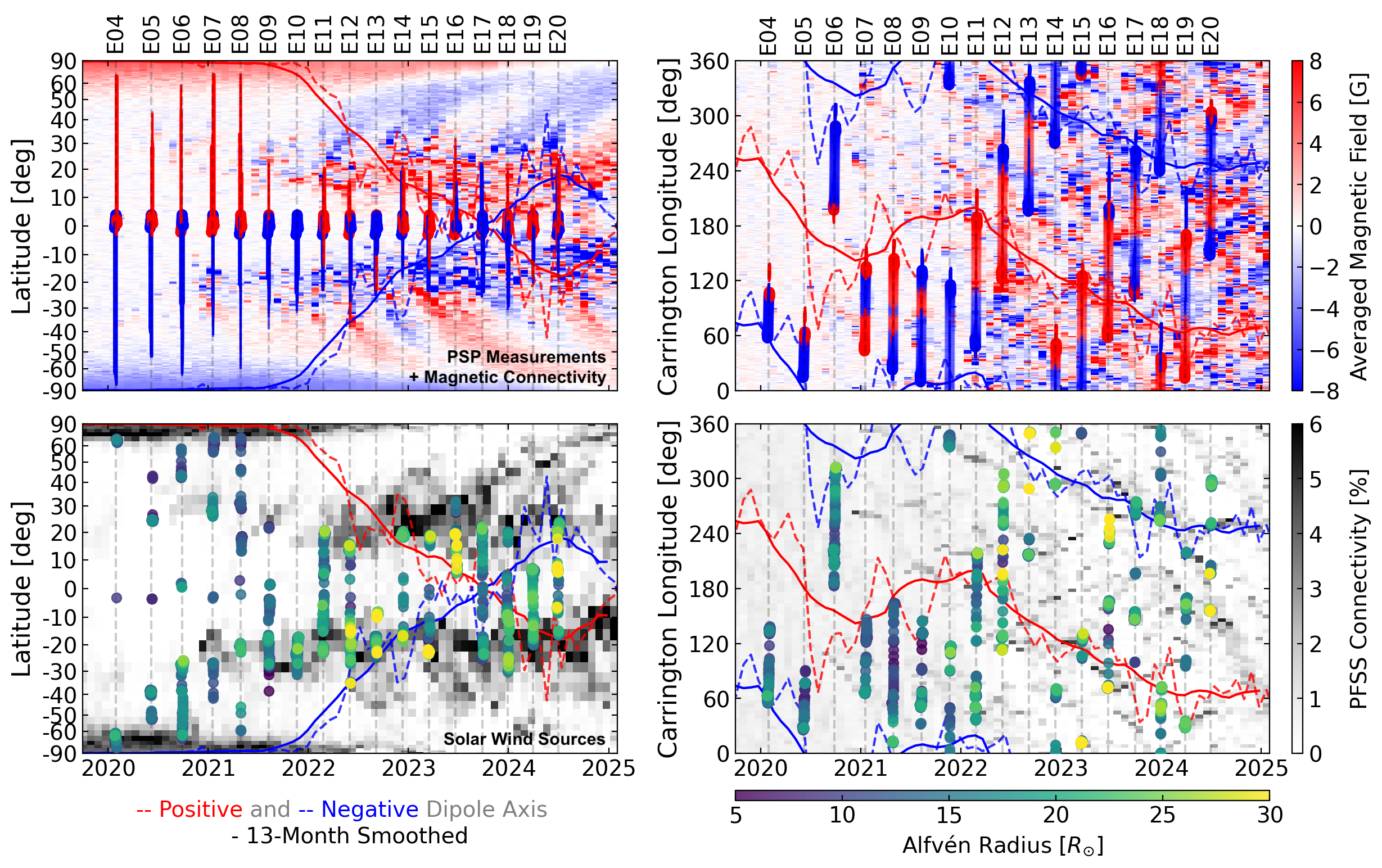}
    \caption{Evolution of PSP's magnetic connectivity and Alfvén radii over solar cycle 25. Top left: The background shows the longitudinally averaged magnetic field of the Sun (magnetic butterfly diagram) with dashed (monthly) and solid (monthly smoothed) lines following the reversal of the dipolar component of the Sun's magnetic field. The magnetic field polarity measured by PSP (markers) and its connectivity in latitude (vertical field lines) from PFSS modelling are shown for each encounter. Bottom left: The background shows the fraction of open magnetic field lines at the solar surface from PFSS modelling with a fixed source surface at 2.5$R_{\odot}$. The reversal of the dipole is shown as above. Markers show the sources of the solar wind (from the field lines in the upper left panel) and are coloured by the size of the Alfvén surface. Right panels: Same as the left panels, but now showing structure in Carrington longitude using latitudinally averaged data.}
    \label{fig:pfss_latlon}
\end{figure*}

\section{Spatial versus temporal fluctuations in the size of the Alfvén surface}

To distinguish the spatial and temporal evolution of the Alfvén surface, we split each encounter of PSP into a longitudinal scan at closest approach, and two radial scans (inbound and outbound), when the tangential speed of the spacecraft is similar to the Sun's rotation rate. We assumed that the longitudinal scans primarily sampled the spatial variation of the Alfvén surface whereas the radial scans were more sensitive to time dependent fluctuations. We used a threshold of the 75th percentile for each encounter on the tangential speed $d\phi/dt$ to identify the longitudinal scans, and on the ratio of the radial speed over the tangential speed $dr/d\phi$ to identify the radial scans (highlighted in Figure \ref{fig:scanning}). Both longitudinal and radial scans had durations between one to three days.  

Figure \ref{fig:scanning} compares the 25th, 50th and 75th percentiles of the hourly Alfvén radii from the longitudinal versus the radial scans in each encounter. The longitudinal and radial scans had a similar distribution of values, except for E12, E13, E15, E18 and E20. These encounters contained a differing proportion of heliospheric current sheet and pseudostreamer crossings in either the longitudinal or radial scans; discussed in Section 5 (see also Appendix \ref{ap:summary} Figure \ref{fig:massmagsummary}). For each encounter, we used discrete Fourier transforms to decompose the fluctuations in the Alfvén surface into spatial scales in the longitudinal scan, and time scales in the radial scans. Figure \ref{fig:scanning} shows the evolution of these spectra and their average. Fluctuations on the order of one solar radii and above were found at scales larger than 15$^{\circ}$ Carrington longitude, and longer than 7 hours in the radial scans. The encounters with the largest average Alfvén radii (like E13 and E15) had the strongest fluctuations at scales greater than 60$^{\circ}$ longitude, which are comparable with the longitudinal scans themselves. A break in the longitudinal scales at 15-20$^{\circ}$ separates power-law slopes of -0.8 and -1.9, whereas the time scales are represented by a slope of -1.2. The inbound and outbound radial scans are treated individually, and so the range of scales was reduced in comparison to the longitudinal scans. Large-scale fluctuations were found to be 10-20\% of the average Alfvén radius at solar minimum and maximum, but up to 40\% during the rising phase of activity (E12 to E15). Nine of the encounters contained longitudinal structure at $\sim$10-20$^{\circ}$ that could be connected to the formation of the solar wind, like switchback patches \citep[e.g.][]{fargette2021characteristic}.

\section{Connecting the solar corona to the Alfvén surface}

To explore the role of the evolving coronal magnetic field on the Alfvén surface, we mapped the in-situ measurements down to the solar surface using a combination of Parker spiral and potential field source surface (PFSS) modelling \citep{altschuler1969magnetic, schrijver2003photospheric}. The PFSS modelling followed the methodology described in \citet{finley2023accounting}. For each encounter, we used the polarity of the interplanetary magnetic field measured by PSP, Solar Orbiter \citep{horbury2020solar} and the Wind spacecraft \citep{lepping1995wind} along with the structure of the solar corona in scattered white light observations from SOHO/LASCO-C2 \citep{domingo1995soho} to select an optimised GONG-ADAPT\footnote{Data accessed May 2025: https://gong.nso.edu/adapt/maps/gong} magnetogram and source surface radius for each PFSS model. Appendix \ref{ap:pfss} describes this process in more detail. Figure \ref{fig:scanEncounter} shows an example of tracing the solar wind from PSP down to the solar surface in E12 (referred to as magnetic connectivity).

Figure \ref{fig:pfss_latlon} shows the observations from E04 to E20 in context with the evolution of the Sun's photospheric and coronal magnetic field. As solar activity increased, the Sun's magnetic field shifted from an axisymmetric dipole at solar minimum to a complex multipolar configuration at solar maximum \citep{derosa2012solar, finley2023evolution}. The dipole's influence extends out into the heliosphere and typically structures the solar corona and wind at large-scales \citep[][]{usmanov2000global, reville2015effect}. The dipole component of the photospheric magnetic field was extracted from SDO/HMI Carrington magnetograms \citep{scherrer2012helioseismic}, as in \citet{finley2024nested}. The polarity of the magnetic field measured by PSP is seen to be correlated with the dipolar magnetic field at large-scales in Figure \ref{fig:pfss_latlon}. The dipole component reversed around 3.5 years into the solar cycle which influenced the magnetic connectivity of PSP. At solar minimum, PSP measured the solar wind that emerged from near to the Sun's polar regions and, due to its orbit at the time, sampled a small range of Carrington longitudes. As the solar cycle progressed, PSP more frequently connected to source regions around the active latitudes and sampled a broader range of longitudes. Fortuitously, E07 to E11 sampled a similar range of Carrington longitudes; while solar activity was increasing. Despite some similarities, discussed in Appendix \ref{ap:comp}, the time-evolution of the coronal magnetic field was sufficient to ensure that each encounter was mostly connected to different regions on the solar surface.

In global solar wind simulations, the Alfvén surface is largest along the dipole axis, and pinches inward over current sheets and pseudostreamers \citep{usmanov2003tilted, pinto2011coupling, reville2017global}. In addition, the rapid expansion of the coronal magnetic field near coronal hole boundaries and above active regions has been shown to form low-Mach boundary layers that locally increase the size of the Alfvén surface \citep[see][]{liu2023generation, jiao2023properties}. The expansion of the Alfvén surface in 2022 coincided with a shift in the inclination of the Sun's dipolar magnetic field away from the rotation axis. E12 had the second largest longitudinally averaged Alfvén radii from E04 to E20, surpassing all of the averaged values from the encounters during the maximum of activity (E16 to E20). Figure \ref{fig:scanEncounter} shows the magnetic connectivity of PSP during E12. The Sun's dipole magnetic field was almost fully inclined to the rotation axis (see Figure \ref{fig:pfss_latlon}) and solar activity was dominated by a nested active region in the northern hemisphere \citep[see][]{finley2025prolific}. The extent of the Alfvén surface was modulated by structure in the underlying coronal magnetic field, with the largest decrease located within 20$^{\circ}$ of the heliospheric current sheet. Similar decreases were found near the pseudostreamers, where the magnetic connectivity jumps without a change in magnetic field polarity \citep{wang2012nature}. At its closest approach in E12, PSP scanned across 60$^{\circ}$ in Carrington longitude of solar wind connected to sources near the positive dipole axis (more open magnetic flux), with a rapid expansion in the low corona that decreased the solar wind density and boosted the Alfvén speed. Thus, PSP is likely to have measured the most extended part of the Alfvén surface at that time. 

In Figure \ref{fig:pfss_latlon}, the largest hourly Alfvén radii were traced back to solar wind sources around the active latitudes, and notably during times when the dipole axis crossed those latitudes. The early encounters, E04 to E09, measured solar wind from higher latitudes and shifted away from the dipole axis in Carrington longitude. E12 was the first encounter to be aligned with the latitude and longitude of the dipole axis (in this case, the positive polarity), and measured an increase in the solar wind magnetic flux. On the other hand, E17 had a much smaller average Alfvén radius (13 solar radii) than similar encounters near solar maximum (15-16 solar radii) as a result of PSP spending a large fraction of the encounter in proximity to the heliospheric current sheet and pseudostreamers. The most recent encounters (E17 to E20) sampled a broad range of Carrington longitudes (up to 180$^{\circ}$) with a complex coronal magnetic field, and so their average Alfvén radii may be more representative of the equatorial solar wind (although E17 is a clear caveat to this).

\section{Solar wind angular momentum-loss}

The size of the Alfvén radius controls the efficiency of the Sun's wind braking and subsequent spin-down during the main sequence \citep{finley2019solar}. Simulations of magnetised stellar winds have been used to parametrise the wind braking torque for Sun-like stars over a wide range of parameters that includes different coronal magnetic field geometries \citep{reville2015effect}, wind heating \citep{pantolmos2017magnetic}, acceleration mechanisms \citep{hazra2021modeling}, and differential rotation \citep{ireland2022effect}. Mass and magnetic flux conservation along a steady flux tube can be used to show the following dependence of the average Alfvén radius $\langle r_A\rangle$ on the wind magnetisation parameter $\Upsilon_{\text{open}}$,
\begin{equation}
    \frac{\langle r_A\rangle}{R_{\odot}}= K \Upsilon_{\text{open}}^m = K\bigg[\frac{\phi_{\text{open}}^2/R_{\odot}^2}{\dot{M} v_{\text{esc}}}\bigg]^m,
    \label{ra_scaling}
\end{equation}
where the Sun's radius is $R_{\odot}=6.96\times 10^{10}$ cm, the Sun's escape speed is $v_{\text{esc}}=618$ km/s, and the Sun's mass-loss rate and open magnetic flux were approximated with $\dot{M} = 4\pi \langle F_m \rangle$ and $\phi_{\text{open}} = 4\pi \langle |F_B| \rangle$, respectively. The parameters $K$ and $m$ account for the multi-dimensional nature of the stellar wind outflow and were fit to the various 2.5D wind simulations \citep[e.g. $K=0.46$ and $m=0.33$ from][]{finley2018dipquadoct}.

\begin{figure*}
    \centering
    \includegraphics[trim=0cm 0cm 0cm 0cm, clip, width=\textwidth]{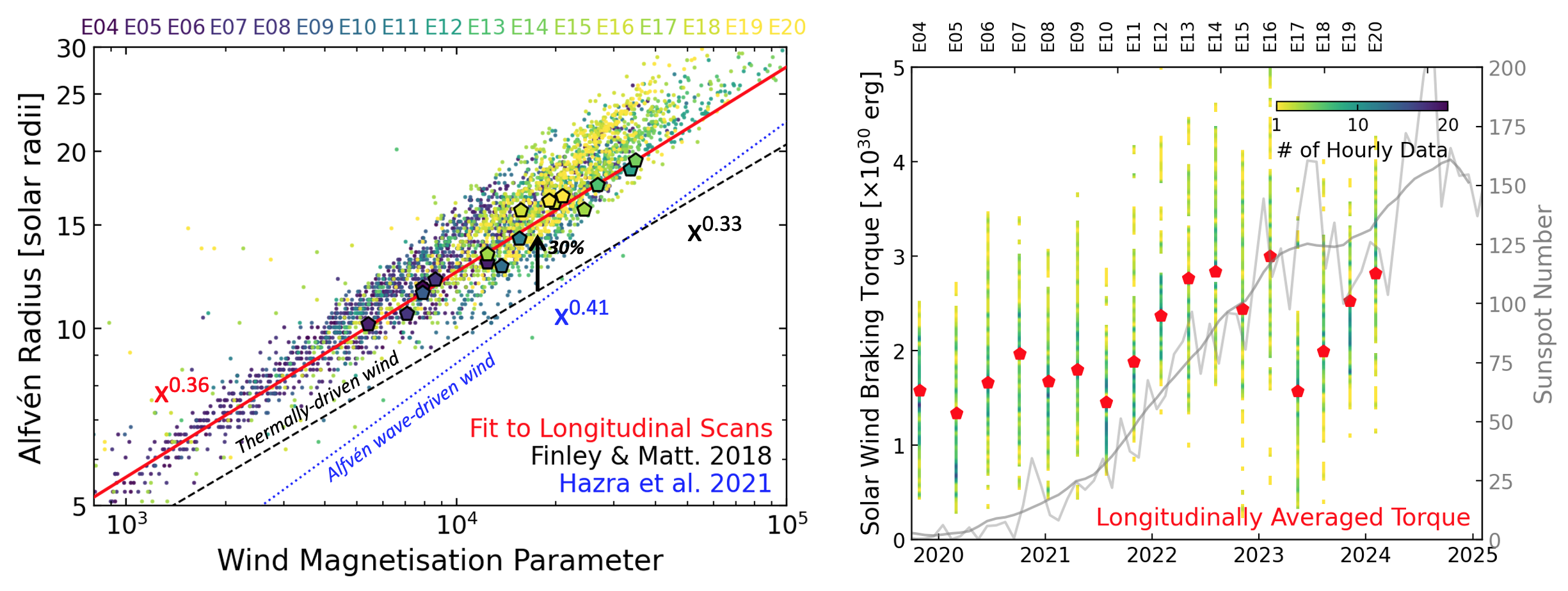}
    \caption{Dependence of the Alfvén radius on the wind magnetisation, and the time-evolution of the Sun's wind braking torque. Left: The longitudinally averaged Alfvén radii are shown with large markers coloured by the monthly sunspot number. The solid red line is fit to those points. The dashed and dotted lines compare the predicted scaling based on 2.5D wind simulations from \citet{finley2018dipquadoct} and \citet{hazra2021modeling}, respectively. The hourly Alfvén radii from each encounter are scattered in the background with the corresponding colour. Right: Same as Figure \ref{fig:cycle}, but now for the solar wind angular momentum-loss rate.}
    \label{fig:torquecycle}
\end{figure*}

We compared the Alfvén radii from PSP with the wind magnetisation parameter in Figure \ref{fig:torquecycle}. Using the longitudinally averaged values of the Alfvén radius, mass-loss rate, and open magnetic flux (to compute $\Upsilon_{\text{open}}$), we found a similar dependence of $m=0.36$ to the 2.5D wind simulations but with the Alfvén radii $\sim30\%$ larger. The same was true when comparing either thermally-driven or Alfvén wave-driven winds. We show the hourly Alfvén radii versus $\Upsilon_{\text{open}}$ in the background with a similar trend and a larger scatter. This scatter was due to a combination of uncertainty in the global mass-loss rates and open magnetic fluxes (extrapolated from local hourly averaged values) and uncertainty in the radial distance of the Alfvén surface when PSP was further away from the Sun. The vertical scatter in the hourly points is therefore correlated with the radial distance of PSP. Despite the scatter, there is a strong correlation of the Alfvén radius with wind magnetisation, especially when using the longitudinally averaged values. The offset of $\sim30\%$ from the 2.5D wind simulations may have a few contributing factors. Firstly, 3D effects due to the non-axisymmetry of the Sun's large-scale magnetic field are not be fully encapsulated. Similarly, the simulations were steady-state, whereas the solar wind is inherently time-dependent which could have influenced our assessment of the wind magnetisation parameter. Equation (\ref{ra_scaling}) is also dependent on the wind acceleration, which could in future be accounted for by correcting the wind magnetisation parameter using the wind speed at the Alfvén surface $v(r_A)$ in place of $v_{\text{esc}}$ \citep[see][]{pantolmos2017magnetic}. Turbulence has also been shown to play a role in reducing the angular momentum-loss rate \citep[by 3-10\%, see][]{chhiber2025effect} and was not accounted for in torque-averaged Alfvén radii from the wind simulations. Finally, both the average Alfvén radius and the wind magnetisation are global parameters and so measurements of the solar wind away from the ecliptic plane may be necessary to make a fair comparison.

Assuming a spherically symmetric outflow, we estimated the Sun's wind braking torque as
\begin{equation}
    \tau = \frac{2}{3}\dot{M}\Omega_{\odot}\langle r_A \rangle^2,
\end{equation}
where the longitudinally averaged mass-flux and Alfvén radius were assumed to be representative of the Sun's mass-loss rate $\dot{M}$ and averaged Alfvén radius $\langle r_A\rangle$ \citep{weber1967angular}. Figure \ref{fig:torquecycle} shows the evolution of the Sun's angular momentum-loss rate as solar activity increased. As the mass-loss rate (see Figure \ref{fig:massmassgcycle}) decreased slightly, the evolution of the wind braking torque followed a weaker trend than the average Alfvén radii in Figure \ref{fig:cycle}. The estimated wind braking torque was highly dependent on the conditions experienced by PSP in each encounter. Overall, the angular momentum-loss rate increased as a function of solar activity from 1-3$\times 10^{30}$ erg. This matches the solar cycle variation in the solar wind angular momentum flux found at 1au with the Wind spacecraft \citep{finley2019direct} and the magnitude of angular momentum-loss from previous studies with PSP \citep{finley2020solar} and Solar Orbiter \citep{verscharen2021angular}. The strength of the Sun's wind braking torque remains at least a factor of two smaller than the value required for the Sun to follow the \citet{skumanich1972time} spin-down, which supports the weakened magnetic braking hypothesis for Sun-like stars around the solar age \citep{van2016weakened}. The application of scaling relations like equation (\ref{ra_scaling}) to other Sun-like stars with observed magnetic fields recovered a similar decrease in wind braking \citep[e.g.][]{metcalfe2022origin}. However, the Sun remains the only star for which these scaling relations can be directly tested.

\section{Conclusions}

We used hourly averaged data from PSP (E04 to E20) to recover the complex, time-varying shape of the Alfvén surface as solar activity increased in solar cycle 25. We found that the longitudinally averaged Alfvén surface expanded from 11 solar radii at solar minimum to 16 solar radii near solar maximum. The dipole component of the Sun's magnetic field became highly inclined after 2022 and the sources of the solar wind measured by PSP shifted towards the Sun's active latitudes. Direct comparison of the average Alfvén radius between encounters was complicated by structure in the coronal magnetic field modulating the solar wind magnetic flux and the local extent of the Alfvén surface. Crossing the Alfvén surface above the dipole axis, or areas of rapidly expanding coronal magnetic field, locally increased the Alfvén radius whereas crossing the heliospheric current sheet or pseudostreamers caused a decrease \citep[as expected from global solar wind simulations, e.g.][]{keppens1999numerical, usmanov2003tilted}.

Using the radial and longitudinal scans of the solar wind from PSP, we found that the Alfvén surface fluctuates by 10-40\% from the average Alfvén radius in both time and space. The strongest longitudinal variation was observed in the rising phase of activity (E12 to E15). The most significant fluctuations, i.e. bigger than one solar radii, were at scales larger than 15$^{\circ}$ Carrington longitude, and longer than 7 hours in the radial scans. We did not remove transient phenomena, such as coronal mass ejections, from our analysis \citep[e.g.][]{mccomas2023parker}. However, the timescales for these events crossing PSP are on the order of hours and the structure in the Alfvén surface was modulated on timescales of days in both the spatial and temporal domains. The large-scale fluctuations in the Alfvén surface could be connected to the highly dynamic nature of the solar wind escaping the solar corona \citep{deforest2014inbound}. This work supports previous studies that suggested a wrinkled or corrugated Alfvén surface \citep{badman2023prediction, liu2023generation, jiao2023properties}. 

The solar wind angular momentum-loss rate using the average Alfvén radii and mass-loss rates from PSP increased from $\sim 1\times 10^{30}$ erg at solar minimum to $\sim 3\times 10^{30}$ erg at solar maximum. This range of values was consistent with previous in-situ measurements. We found that the Alfvén radii from PSP were $\sim30\%$ larger than predicted by global wind simulations with the same mass-loss rate and open magnetic flux. This could be due to a combination of 3D effects (just inclining the dipole component in 3D can increase the average Alfvén radius by 10\%), time-dependent effects, turbulence, and the uncertainty in extrapolating the equatorial mass and magnetic flux into a global mass-loss and open magnetic flux. Despite this, the average Alfvén radii from PSP had the same dependence on the wind magnetisation parameter (ratio of the open magnetic flux and mass-loss rate) as both thermally-driven and more realistic Alfvén wave-driven winds.

Using PSP, we explored the structure of the Alfvén surface around the solar equator. In future, in-situ measurements of the solar wind from outside of the ecliptic plane by ESA's Solar Orbiter \citep{muller2020solar, liu2023generation} could provide an opportunity to better constrain the latitudinal variation of the Alfvén surface and the Sun's angular momentum-loss rate. After the launch of NASA's Polarimeter to Unify the Corona and Heliosphere (PUNCH) mission \citep{deforest2022polarimeter}, we anticipate more remote-sensing observations capable of constraining the large-scale morphology of the Sun's Alfvén surface and mass-loss rate.

\begin{acknowledgements}
This research was inspired by the ``Star-Planet Interaction and Aeronomy. Let's Talk About Physics'' workshop that was held in November 2024 in Saint Malo, France. We thank the organisers, Antoine Strugarek and Antonio García Muñoz, as well as all the participants that contributed to the engaging scientific discussions. 

This research has received funding from the European Research Council (ERC) under the European Union’s Horizon 2020 research and innovation programme (grant agreement No 810218 WHOLESUN), in addition to funding by the Centre National d'Etudes Spatiales (CNES) Solar Orbiter, and the Institut National des Sciences de l'Univers (INSU) via the Programme National Soleil-Terre (PNST).

We acknowledge the NASA Parker Solar Probe Mission and particularly the FIELDS team led by S. D. Bale and the SWEAP team led by J. Kasper for use of data. The data used in this study are available at the PSP science gateway https://sppgway.jhuapl.edu/. 

Data supplied courtesy of the SDO/HMI and SDO/AIA consortia. SDO is the first mission to be launched for NASA's Living With a Star (LWS) Program.

The sunspot number used in this work are from WDC-SILSO, Royal Observatory of Belgium, Brussels. 

Data manipulation was performed using the numpy \citep{2020NumPy-Array}, scipy \citep{2020SciPy-NMeth}, and pySHTOOLS \citep{wieczorek2018shtools} python packages.
Figures in this work are produced using the python package matplotlib \citep{hunter2007matplotlib}.
\end{acknowledgements}

%
%

\bibliographystyle{yahapj}
\bibliography{adam}

\onecolumn
\begin{appendix}

\section{Mass and magnetic flux summary}\label{ap:summary}

\begin{figure*}[h!]
    \centering
    \includegraphics[trim=0cm 0cm 0cm 0cm, clip, width=\textwidth]{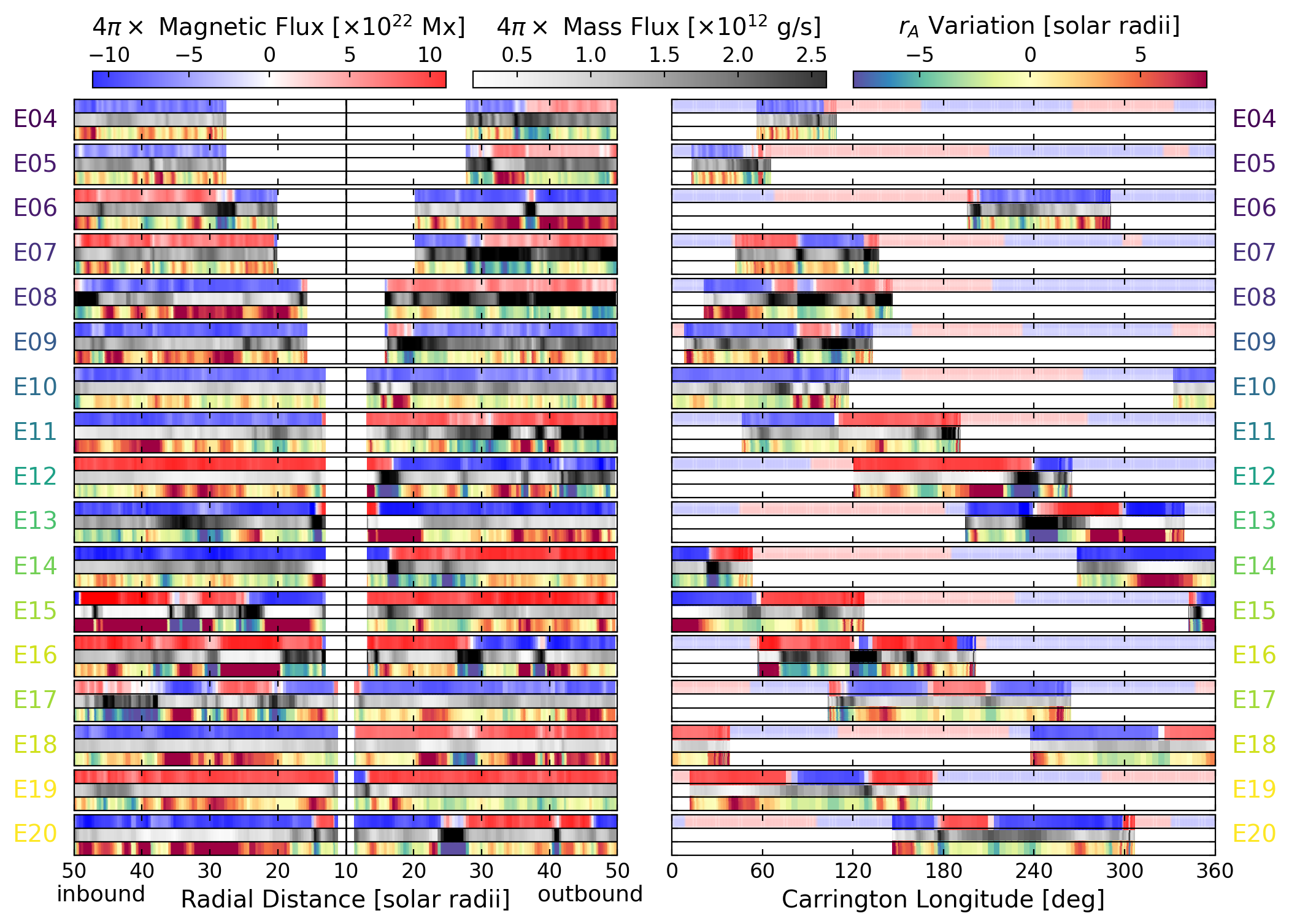}
    \caption{Similar to Figure \ref{fig:speedmachsummary}, but now for the magnetic flux $F_B$, mass flux $F_m$, and variation in the Alfvén radius $r_A-\langle r_A \rangle$. The encounter labels are coloured by solar activity from dark to light. The magnetic field polarity at the solar equator from the PFSS models is shown with light blue and red in the background of the magnetic flux panels.}
    \label{fig:massmagsummary}
\end{figure*}

The Alfvén radii extrapolated from PSP were dependent on the in-situ measurements of the magnetic flux and magnetic flux. Figure \ref{fig:massmagsummary} summarises these measurements from E04 to E20, and compares them to the variation of the Alfvén radius across each encounter. Variation in the Alfvén radius was correlated with the solar wind mass flux. Strong decreases in the mass flux generally correspond to over-expanded, low density, solar wind streams with high Alfvén speeds that caused the Alfvén surface to expand. Conversely, the high solar wind mass flux found near the heliospheric current sheet (where the magnetic flux changes sign) and pseudostreamers decreased the local Alfvén speed and lowers the height of the Alfvén surface. The distribution of mass flux and magnetic flux encountered by PSP evolved as a function of solar cycle (see Figure \ref{fig:massmassgcycle}). From E04 to E09, the solar wind mass flux was systematically larger (40\% of E10 to E20) and the magnetic flux was lower (nearly half). This was consistent with PSP sampling the equatorial solar wind at solar minimum, where the heliospheric current sheet was confined to the equator by the Sun's dipolar magnetic field. From E12 to E16, there was a distinct shift in the magnetic flux towards larger values and the mass flux to smaller values. This was more consistent with PSP sampling the faster and less dense solar wind away from the heliospheric current sheet, like that which escapes from equatorial coronal holes. The magnetic flux in E17 and E18 dropped back down to values more similar with solar minimum, which matches the increased time spent near current sheets and pseudostreamers. The mass and magnetic flux increased in E19 and E20 to values more similar with E12 to E16, but slightly smaller due to the increased number of current sheet crossings found in the longitudinal scans.

\section{PFSS optimisation}\label{ap:pfss}

\begin{figure*}[h!]
    \centering
    \includegraphics[trim=0cm 0cm 0cm 0cm, clip, width=\textwidth]{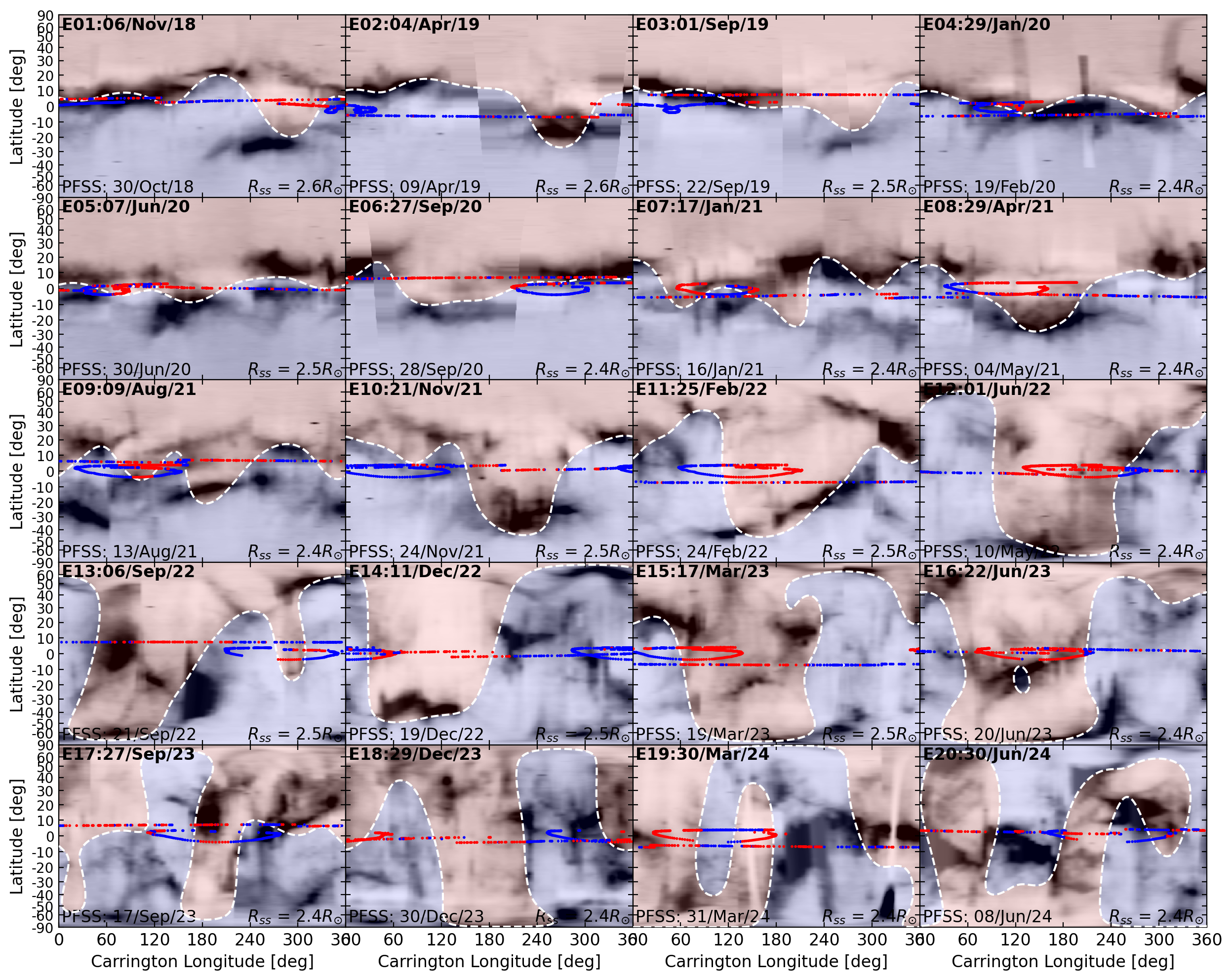}
    \caption{Summary of PFSS models that were fit to the interplanetary magnetic field polarity and scattered white light observations. Coloured markers show the in-situ measurements from PSP, Solar Orbiter, and Wind during two weeks around each encounter. The greyscale shows the scattered white light brightness from SOHO/LASCO-C2 at three solar radii. The red and blue background colour shows the polarity predicted by the PFSS model, with the source surface radius and ADAPT-GONG map date annotated.}
    \label{fig:pfss_opt}
\end{figure*}

\begin{figure*}[h!]
    \centering
    \includegraphics[trim=0cm 0cm 0cm 0cm, clip, width=\textwidth]{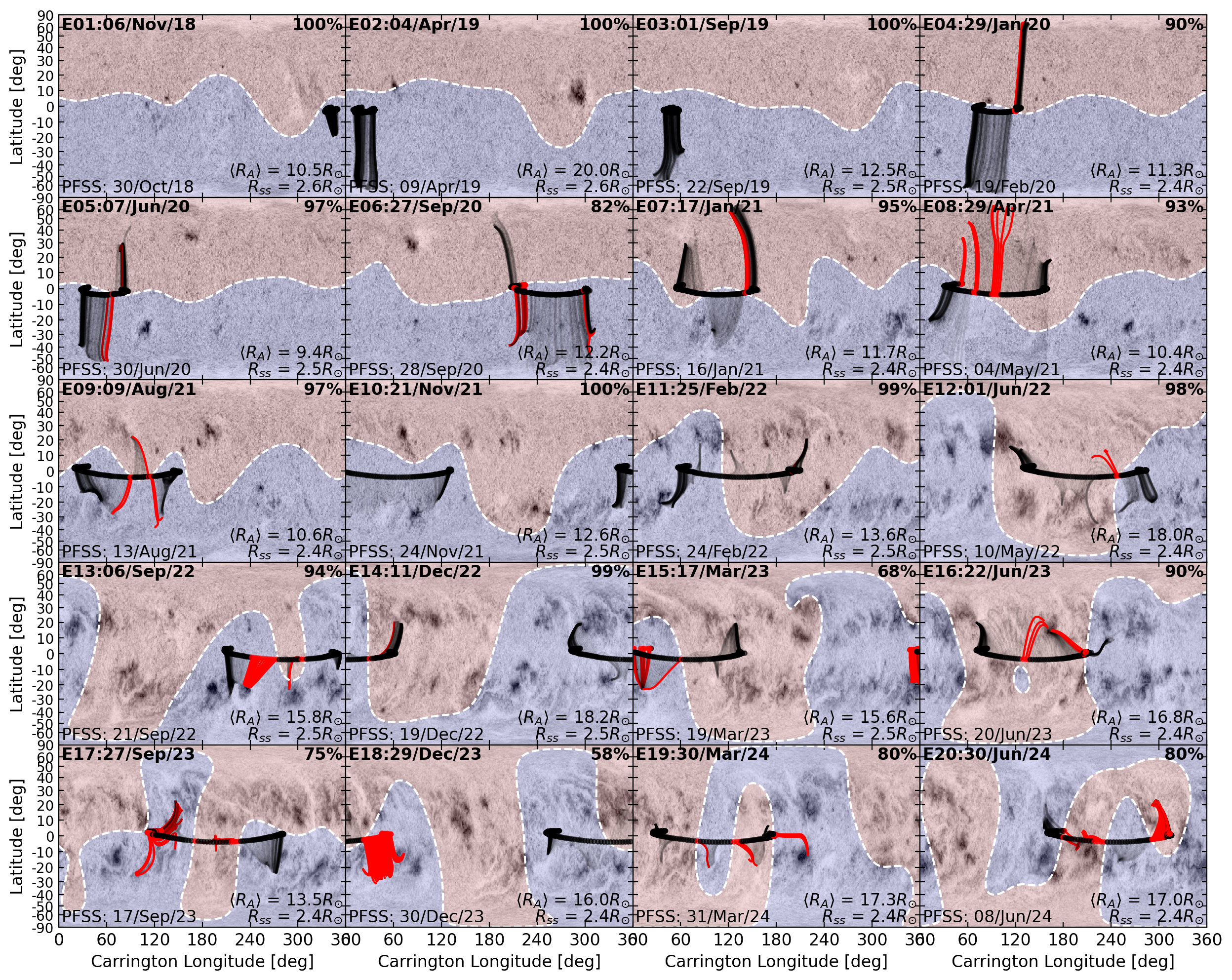}
    \caption{Similar to Figure \ref{fig:pfss_opt}, but now showing the PFSS magnetic field lines from PSP to the solar surface coloured red when the predicted magnetic field polarity does not match the in-situ measurement. The greyscale shows Carrington maps of SDO/AIA 304$\AA$ emission. The fraction of hourly magnetic field polarity that matched the PFSS model is given in the upper right corner for each encounter.}
    \label{fig:pfss_fraction}
\end{figure*}

Placing the in-situ measurements in context with the Sun's dynamic magnetic field is challenging \citep[e.g.][]{badman2020magnetic}. In this study, we opted to use one PFSS model of the coronal magnetic field for each encounter. This was done for simplicity and could be greatly improved in future works by considering the time-varying nature of the solar corona \citep[][]{owens2024importance, downs2025near}. We used ADAPT-GONG Carrington magnetograms as the input for the PFSS modelling \citep{hickmann2015data}. The PFSS model for each encounter was optimised with two free parameters, the date/time of the magnetogram and the source surface radius. The date of the magnetogram was limited to be within plus or minus one solar rotation of the closest approach of PSP. This was done to capture magnetic structures in the PFSS that were observed in-situ, but had either yet to be inserted into the ADAPT-GONG magnetograms or had diffused too quickly. The source surface was allowed to vary between 1.8 and 3.0, however we found that values between 2.4 and 2.6 were favoured.

Each PFSS model was optimised starting with the hourly in-situ magnetic field polarity from PSP, Solar Orbiter and Wind. We selected in-situ data from two weeks before and after each encounter. We used the hourly averaged in-situ solar wind speed to map the observations back to three solar radii (to later compare with coronal observations), using a fixed wind acceleration of $\alpha=0.1$. The in-situ measurements of magnetic field polarity are shown in Figure \ref{fig:pfss_opt}. The magnetic field polarity from PFSS models using a range of dates and source surface radii were compared with these measurements. Priority was given to matching the polarity from PSP (i.e. local small-scale structure), with measurements from Solar Orbiter and Wind used to constrain the Sun's large-scale magnetic field. In addition, we used scattered white light observations from SOHO/LASCO-C2 to break the degeneracy between similar PFSS models. We assumed that higher electron densities in the solar wind corresponded with either the heliospheric current sheet or pseudostreamers.

The optimised PFSS models chosen for each encounter are shown in Figure \ref{fig:pfss_opt}, in comparison with the in-situ and scattered white light observations. It would be hard for any single PFSS model to capture all of this information, especially as the input magnetograms are limited by our single vantage point from Earth that cannot capture magnetic flux emergence on the Sun's far-side \citep{perri2024impact}. However, each PFSS model is representative of the average coronal magnetic field at that time. The evolution of the coronal magnetic field over the solar cycle is visible in the chosen PFSS models, with the current sheet and pseudostreamers rapidly spreading to higher latitudes in 2022. Figure \ref{fig:pfss_fraction} summarises the relative success each PFSS model had in reproducing the hourly magnetic field polarity observed by PSP. We identified intervals that the PFSS modelling failed to capture. The fraction of hourly intervals with the correct polarity in each encounter ranged from 80-100\%, with a few exceptions. Notably, E15 and E17 at 68\% and 58\% respectively. In the context of this study, the PFSS modelling was used to compare the in-situ observations from PSP with the global evolution of the solar corona, and so we did not aim to capture every detail in the observations.

\section{Comparison of consecutive PSP encounters}\label{ap:comp}

\begin{figure*}[h!]
    \centering
    \includegraphics[trim=0cm 0cm 0cm 0cm, clip, width=0.9\textwidth]{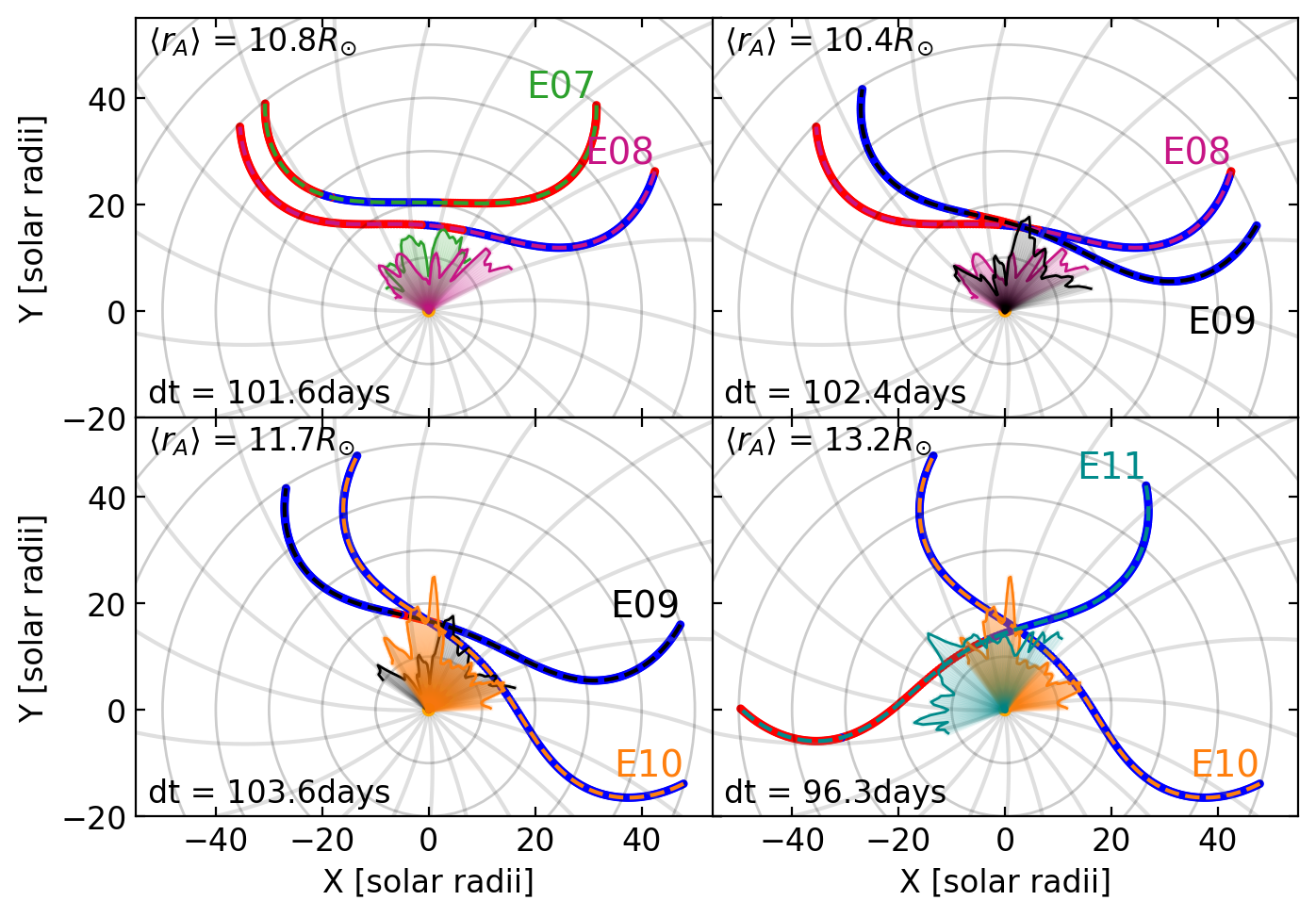}
    \caption{Comparison of the Alfvén surface recovered in consecutive PSP encounters that covered similar Carrington longitudes. From E07 to E11, PSP repeatedly scanned longitudes around 90$^{\circ}$ Carrington longitude. The path of PSP is coloured by the in-situ radial magnetic field polarity. The average Alfvén radius for each combination is given along with the time $dt$ between their respective perihelia.}
    \label{fig:comp1}
\end{figure*}

\begin{figure*}[h!]
    \centering
    \includegraphics[trim=0cm 0cm 0cm 0cm, clip, width=0.3\textwidth]{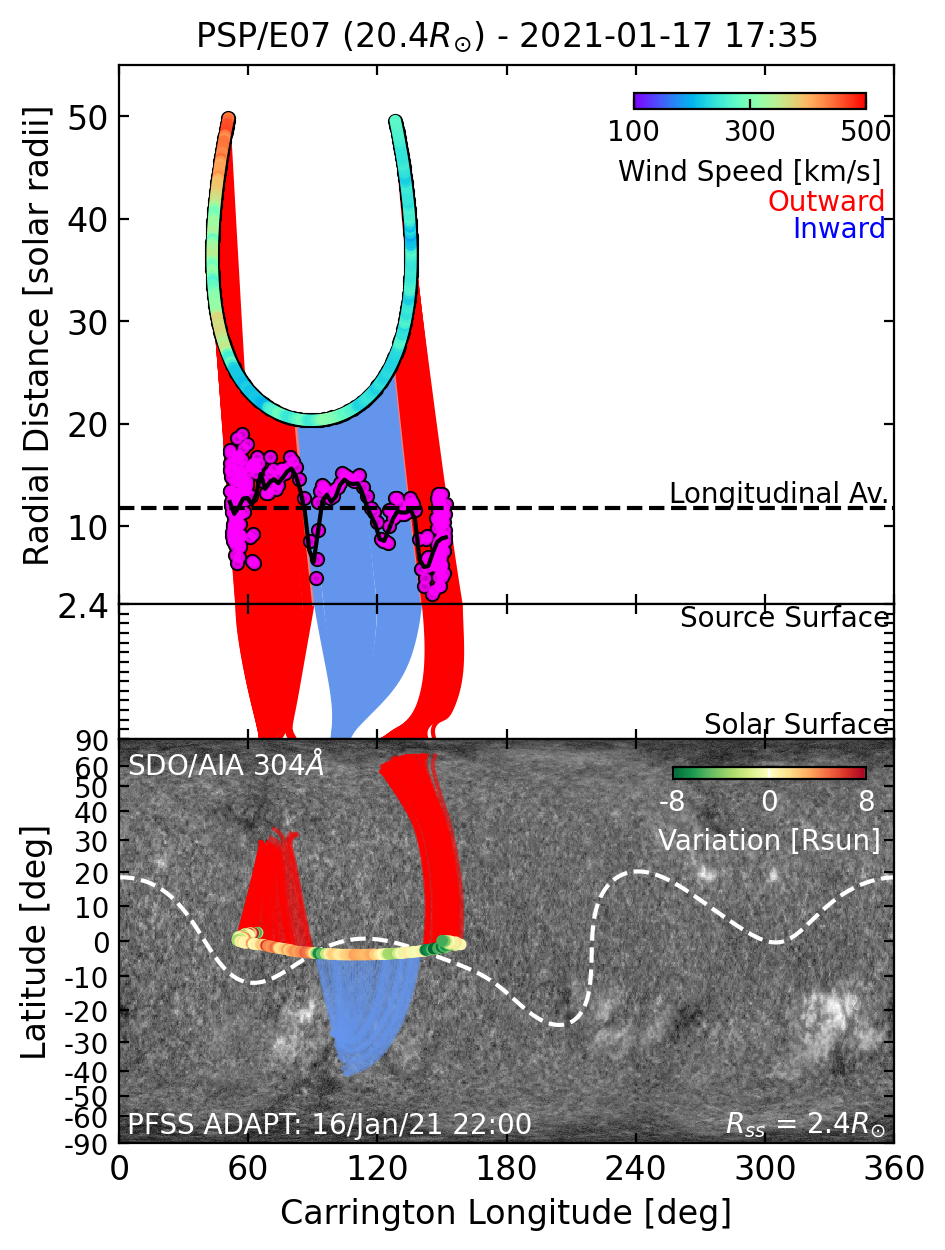}
    \includegraphics[trim=0cm 0cm 0cm 0cm, clip, width=0.3\textwidth]{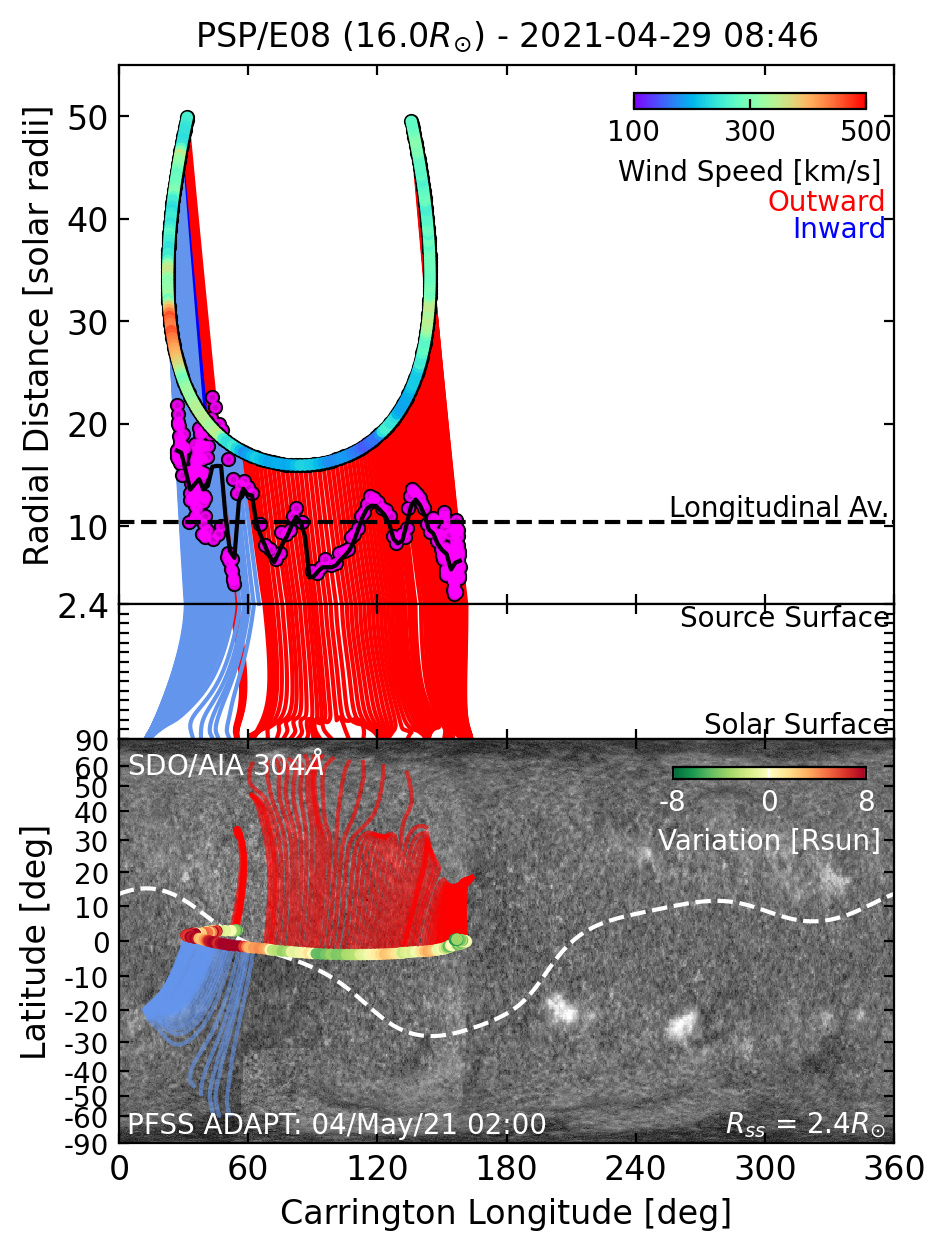}
    \includegraphics[trim=0cm 0cm 0cm 0cm, clip, width=0.3\textwidth]{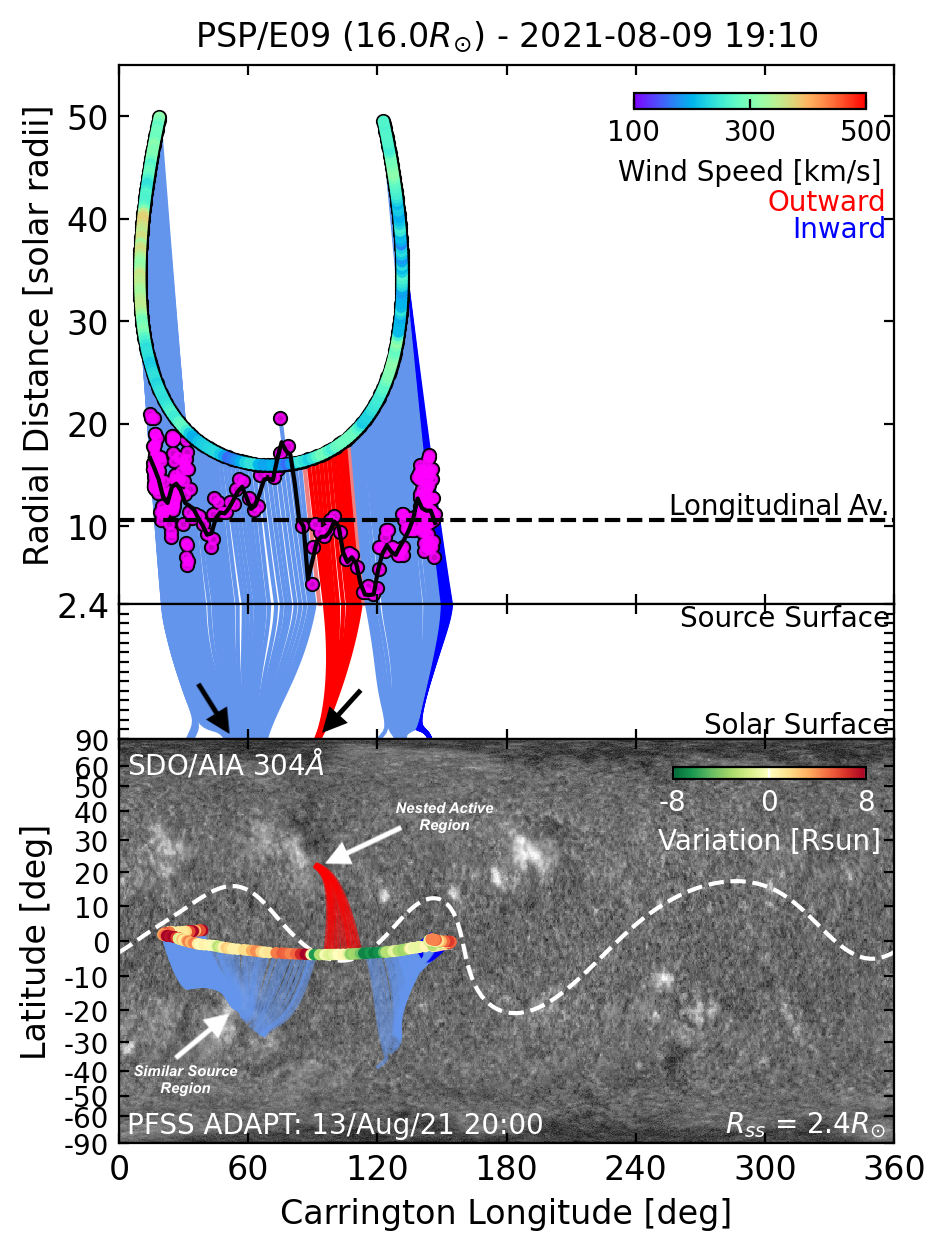}\\
    \includegraphics[trim=0cm 0cm 0cm 0cm, clip, width=0.3\textwidth]{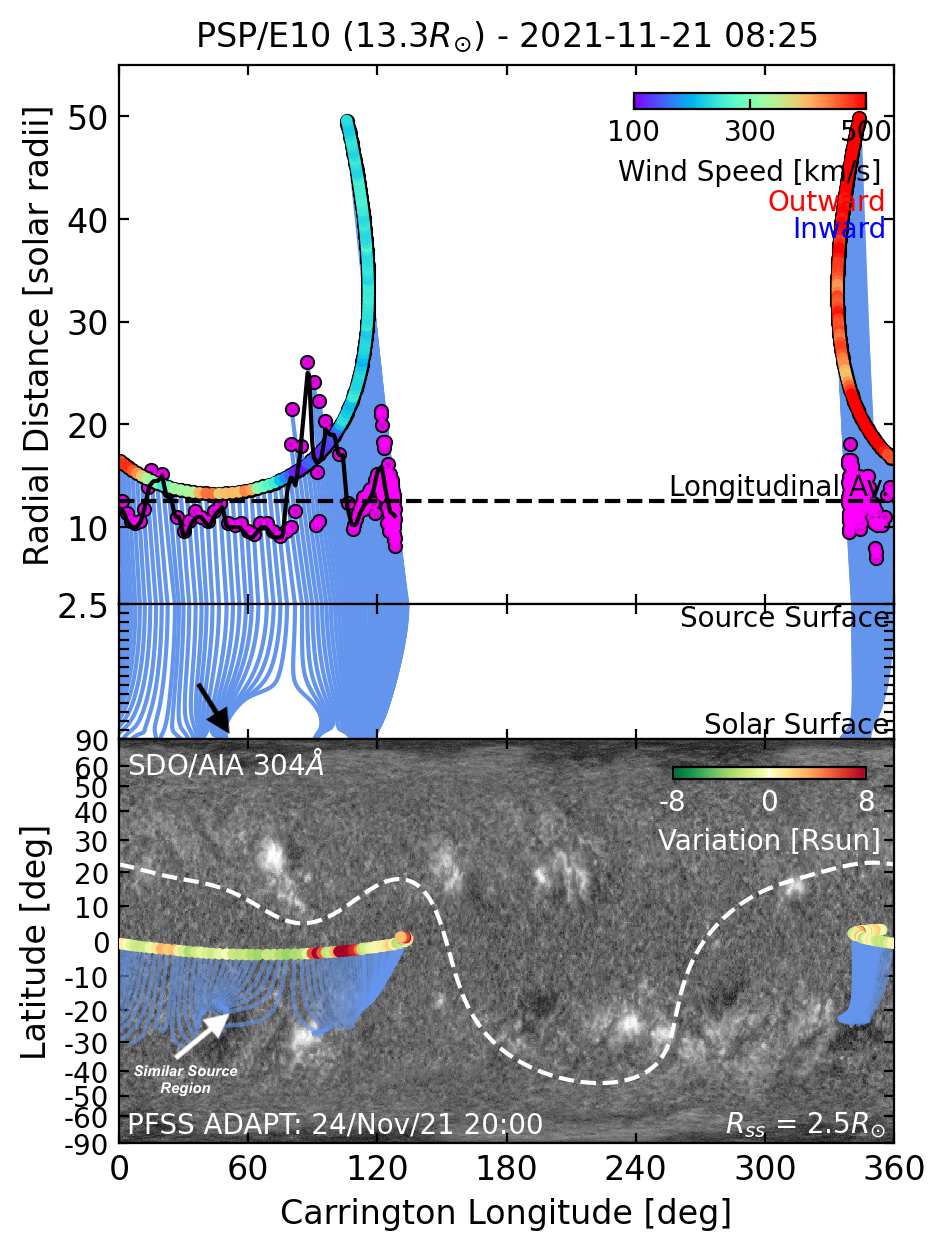}
    \includegraphics[trim=0cm 0cm 0cm 0cm, clip, width=0.3\textwidth]{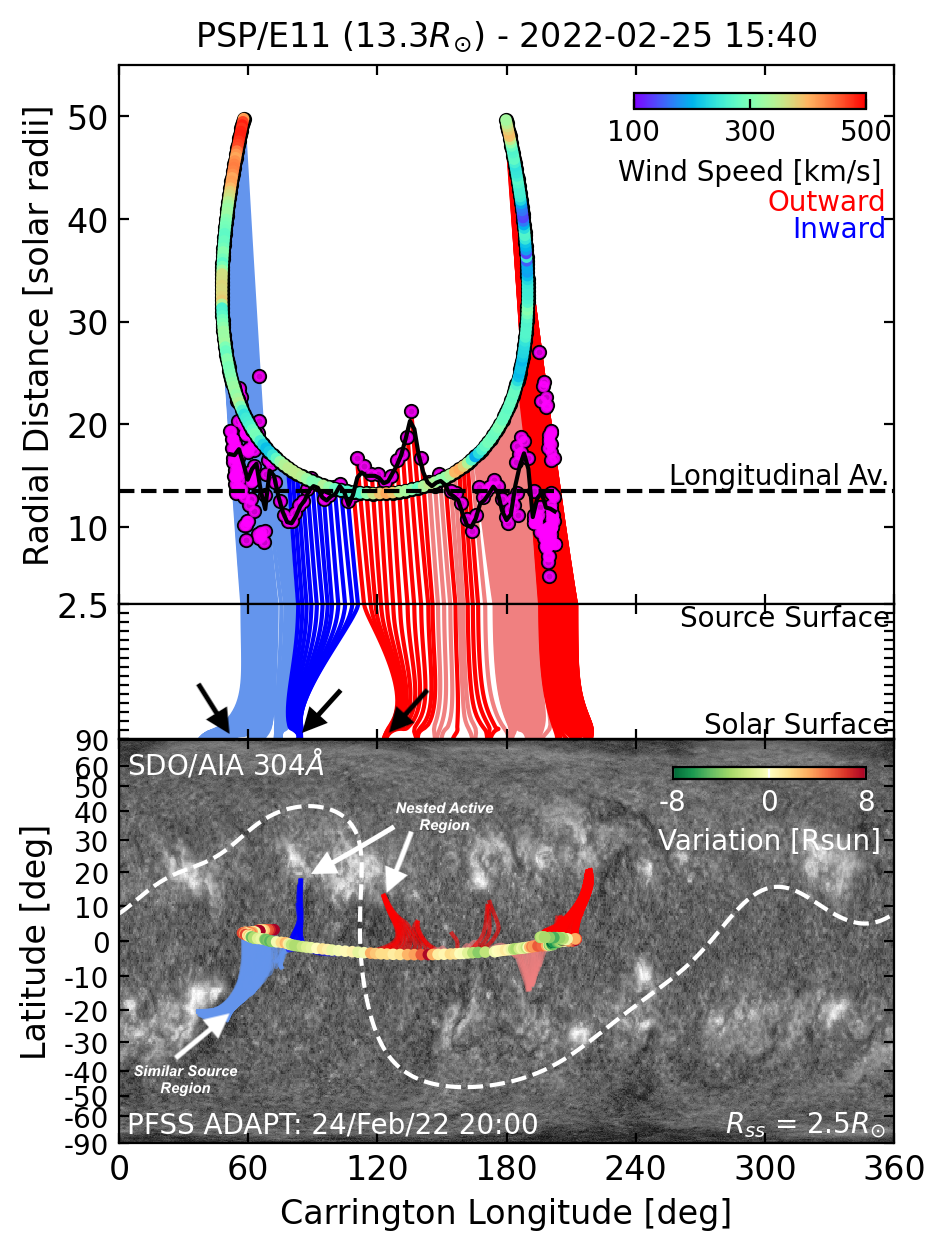}
    \caption{Same as Figure \ref{fig:scanEncounter}, but now for E07 to E11.}
    \label{fig:comp1_pfss}
\end{figure*}

Due to the orbital period of PSP, from E07 to E11 (from January 2021 to February 2022) the spacecraft made its closest approach over a similar range of Carrington longitudes. These perihelia were centered around 90$^{\circ}$ Carrington longitude, and a few degrees south of the solar equator. The interval between each consecutive encounter was around 100 days (3.7 solar rotations). Figure \ref{fig:comp1} compares the extent of the Alfvén surface between each consecutive encounter from E07 to E11. Figure \ref{fig:comp1_pfss}, presents the solar wind back-mapping for each of these encounters; as done for E12 in Figure \ref{fig:scanEncounter}. The size of the Alfvén surface and its fluctuation were comparable between each pair. Despite revisiting similar longitudes, the evolution of the coronal magnetic field meant that each encounter measured very different conditions. From E07 to E11, the average Alfvén radius grew from 11 to 13 solar radii. Fluctuations in the Alfvén surface on the order of 20$^{\circ}$ in longitude were visible, corresponding to the power in Figure \ref{fig:scanning}. The sharpest fluctuation in the Alfvén surface correspond to largest shifts in the magnetic connectivity of PSP (see Figure \ref{fig:comp1_pfss}).

During this sequence, the heliospheric current sheet became inclined to the rotation axis (see also Figure \ref{fig:pfss_opt}), and the solar wind source regions moved toward the active longitudes. This reduced the average number of current sheet crossings, and increased the number of north-south pseudostreamer crossings. E09, E10 and E11 revisited a similar source region around -20$^{\circ}$ latitude and 50$^{\circ}$ Carrington longitude, with three months apart. This area was subject to flux emergence and hosted various active regions. From encounter to encounter, PSP measured the mass and magnetic flux to vary from $0.7-1.4\times 10^{12}$ g/s and $5.7-7.1\times 10^{22}$ Mx (multiplied by a factor of $4\pi$) with the local Alfvén radius varying between $10-18$ solar radii. The largest magnetic flux and Alfvén radius was found in E11 in the presence of an active region. Similarly, E09 and E11 were connected to open magnetic field near to the nested active region that emerged in the northern hemisphere around 90$^{\circ}$ Carrington longitude \citep[see][]{finley2025prolific}, with six months difference. In this case, PSP measured values of $1-2\times 10^{12}$ g/s and $4.6-6.5\times 10^{22}$ Mx with the Alfvén radius varying between $8.8-14$ solar radii, similarly largest when the active nest was stronger in E11.




\end{appendix}

\end{document}